\documentclass[reqno]{amsart}    
\usepackage[english]{babel}
\usepackage{amsmath}
\usepackage{amsthm}
\usepackage{amssymb}
\usepackage{cite}

\newcommand\C{{\mathbb C}}                % complex numbers
\newcommand\N{{\mathbb N}}
\newcommand\R{\mathbb R}

\newcommand\sB{{\stoch B}}

\newcommand\D{{\mathcal D}}

\newcommand\E{{\mathbb E}}
\newcommand\cE{{\mathcal E}}

\newcommand\M{{\mathcal M}}          % manifold

\newcommand\gm{{\/\mathfrak m}}
\newcommand\sM{{\stoch M}} 
       
\renewcommand\L{{\mathcal L}}
\newcommand\cL{{\mathcal L}}

\newcommand\cW{{\mathcal W}}
\newcommand\sW{{\mkern.8mu\stoch W}}    
\newcommand\bP{{\mathbb P}}
\renewcommand\H{{\mathcal H}}
\newcommand\Q{{\mathcal Q}}
\newcommand\cQ{{\mathcal Q}}

\newcommand\cS{{\mathfrak S}}
\renewcommand\t{{\mathcal T}}

\newcommand\sY{{\mkern.5mu\stoch Y}}    
\newcommand\bZ{{\mathbb Z}}
\newcommand\sZ{{\stoch Z}}    

\newcommand\stoch{\mathsf}

\newcommand\Kato{{\mathcal K}(\bP^D)}
\newcommand\Katopm{{\mathcal K_\pm}(\bP^D)}
\newcommand\Kloc{{\mathcal K}_{\mbox{\scriptsize\it loc}}(\bP^D)}

\newcommand\id{\mbox{id}}
\newcommand\Bg{{L^2_{\mbox{\scriptsize\it hol}}(h\meas)}}
\newcommand\hol{{\mbox{\scriptsize\it hol}}}
\newcommand\loc{{\mbox{\scriptsize\it loc}}}
\newcommand\ol{\overline}
\newcommand\Cinfty{\mathop{\hbox{$C$}}^\infty}
\renewcommand\centerdot{\mathbf{\cdot}}

\def\downto{{\mathchoice
{\raise.25ex\hbox{$\,\scriptstyle\searrow\;$}} 
{\raise.25ex\hbox{$\,\scriptstyle\searrow\;$}} 
{\raise.25ex\hbox{$\scriptscriptstyle\searrow$}} 
{\raise.25ex\hbox{$\scriptscriptstyle\searrow$}} 
}}

\newcommand\abs[1]{\left\vert {#1}\right\vert}
\newcommand\ep[1]{{e}^{\textstyle #1}}
\newcommand\norm[1]{{\left\vert\mkern-1.8mu\left\vert#1\right\vert\mkern-1.8mu\right\vert}}

\newcommand\Bnorm[1]{{\Bigl\vert\mkern-2.1mu\Bigl\vert#1\Bigr\vert\mkern-2.1mu\Bigr\vert}}
\newcommand\hnorm[1]{{\biggl\vert\mkern-2.2mu\biggl\vert#1\biggr\vert\mkern-2.2mu\biggr\vert}}
\newcommand\nnorm[1]{{\left\vert\mkern-2mu\left\vert\mkern-2mu\left\vert
                      #1\right\vert\mkern-2mu\right\vert\mkern-2mu\right\vert}}
\newcommand\Cov{\mathrm{Cov}}

\renewcommand\fam[2]{{\{#1\}_{#2}}}
\newcommand\up[1]{^{(#1)}}
\newcommand\grad{\mathop{\mathrm{grad}}}

\newcommand\dz{{d^{\mkern2mu 2n}z}}
\newcommand\qvar[1]{\mbox{\large\textbf{[}}#1\mbox{\large\textbf{]}}}
\newcommand\meas{m}

\newcommand\eq[1]{(\ref{eq:#1})}

\theoremstyle{plain}
\newtheorem{thm}{Theorem}
\newtheorem{dk}[thm]{Theorem (Daubechies-Klauder formula)}
\newtheorem{fk}[thm]{Proposition (Feynman-Kac formula)}
\theoremstyle{definition}
\newtheorem{defn}[thm]{Definition}

\newtheorem{rem}[thm]{Remark}
\newtheorem{remarks}[thm]{Remarks}
\newtheorem{convention}[thm]{Convention}
\newtheorem{conventions}[thm]{Conventions}
\newtheorem{lem}[thm]{Lemma}
\newtheorem{prop}[thm]{Proposition}
\newtheorem{consequence}[thm]{Consequence}

\newtheorem{appdef}{Definition}[section]
\newtheorem{approp}[appdef]{Proposition}
\newtheorem{applem}[appdef]{Lemma}

\newtheorem{appcon}[appdef]{Consequence}
\newtheorem{appthm}[appdef]{Theorem}
\newtheorem{itcomment}[thm]{Comment}
\newtheorem{itconsequence}[thm]{Consequence}

\newtheorem{itremarks}[thm]{Remarks}

\begin{document}

\author{Bernhard G. Bodmann}
\address{Bernhard G. Bodmann, 337 Jadwin Hall, Physics Department, Princeton University,
  Princeton, NJ 08544}
\email{bgb@princeton.edu}
\title[Self-Adjoint Berezin-Toeplitz Operators on K{\"a}hler 
Manifolds]{Construction of Self-Adjoint Berezin-Toeplitz Operators 
on K\"ahler Manifolds and a Probabilistic Representation of the 
Associated Semigroups}

\begin{abstract}
We investigate a class of operators resulting from a
quantization scheme attributed to Berezin.
These so-called Berezin-Toeplitz operators are
defined on a Hilbert space of square-integrable holomorphic 
sections in a line bundle over the classical 
phase space. As a first goal we develop
self-adjointness criteria for Berezin-Toeplitz operators 
defined via quadratic forms. Then, following a
concept of Daubechies and Klauder,
the semigroups generated by these 
operators may under certain conditions be  
represented in the form of Wiener-regularized 
path integrals. 
More explicitly, the integration is 
taken over Brownian-motion
paths in phase space in the 
ultra-diffusive limit.
All results are the consequence of a relation 
between Berezin-Toeplitz 
operators and Schr\"o\-din\-ger operators 
defined by certain quadratic forms. 
The probabilistic representation is derived 
in conjunction with a 
version of the Feynman-Kac formula.
\end{abstract}


\date{\today}

\maketitle

\section{Introduction}

\subsection{Scope of this work}

The general theme in this work is the geometric formulation of 
Berezin-Toeplitz quantization on K\"ahler manifolds. This quantization 
prescription was introduced by Berezin \cite{Ber72a,Ber74} to construct 
quantum models with the help of certain continuous representations in 
the sense of Klauder \cite{Kla63a,Kla63b,Kla64,KM65,KMK65,MK64}, 
more specifically by using spaces of holomorphic functions on phase-space 
manifolds with a K\"ahler structure. Cahen, Gutt,
Rawnsley and others \cite{BMS94,CGR90,CGR93,CGR94,CGR95} subsequently
cast Berezin's construction in a manifestly coordinate-in\-de\-pen\-dent form 
by borrowing ideas from geometric quantization \cite{Kos70,Sni80,Sou66}. 
In this form the quantum kinematics is encoded in a Hilbert space of 
square-integrable, holomorphic sections in a holomorphic line bundle. 
A Berezin-Toeplitz operator $T_f$ on such a Hilbert space is 
characterized by its associated sesquilinear form, which is obtained 
by multiplying the measure in the $L^2$-inner product of the Hilbert space
with a sufficiently regular real-valued function $f$.
The quantization context arises from interpreting
this function as a classical observable that is in
some sense in correspondence with $T_f$. 
Indeed, one may prove that the precise notion
of a correspondence principle applies in the case of homogeneous or compact 
K\"ahler manifolds, see \cite{Ber74,Per86} or \cite{BMS94,Sch98}.

A first goal in this work is to derive conditions for the validity of 
this quantization procedure. More precisely, we obtain regularity conditions 
for possibly unbounded classical Hamiltonians ensuring that their quantum 
analogues are self-adjoint operators. The discussion of these conditions
develops from a rather abstract level to concrete criteria in terms
of the Kato class that is intrinsically determined by the underlying 
geometry. 

The remaining part of this work generalizes an approach to path-integral 
quantization proposed by Daubechies and Klauder 
\cite{DK82,DK85,DK86,KD83,KD84}; see 
also \cite{DKP87}. Superficially, it is a phase-space version of Feynman's 
path integral that has been rendered mathematically well-defined by a 
Wiener-measure regularization. However, a closer look shows that 
the construction by Daubechies and Klauder can be understood as a 
path-integral formulation of Berezin-Toeplitz quantization on certain 
homogeneous K\"ahler manifolds. Indeed, 
a generalization to arbitrary K\"ahler manifolds has been 
advocated in several publications \cite{AK96,AKL93,Kla94,KO89}
and carried out for the compact case by Charles \cite{Cha99}. 
The advocated generalization is a probabilistic expression
for the unitary group $\{ e^{-itT_f} \}_{t \in \R}$ 
generated by a Berezin-Toeplitz operator $T_f$. 
More precisely, a Wiener-regularized 
path integral expresses the integral kernel
of the time-evolution operator $e^{-itT_f}$
as the  ultra-diffusive limit of an expectation value 
over Brownian motion paths on the classical phase space.

In contrast to the setting considered by Charles \cite{Cha99}, we include 
the case of unbounded Berezin-Toeplitz operators
and non-compact manifolds, subject to certain technical conditions. 
Moreover, we show that instead of the Brownian motion governed
by the original K\"ahler metric as in \cite{Cha99}, the Wiener 
regularization may be realized using a conformally
rescaled metric, at the cost of adjusting the path measure with a suitable 
Feynman-Kac functional. A minor difference with the original intent of 
Daubechies and Klauder and its advocated generalizations 
\cite{AK96,AKL93,Kla94,KO89} is that instead of unitary groups, we 
focus on the probabilistic representation 
of semigroups $\{ e^{-tT_f} \}_{t \ge 0}$ that are generated by self-adjoint, 
semibounded Berezin-Toeplitz operators.
The expression for $e^{-tT_f}$ is entirely geometric in 
nature and opens up a wealth of analytic tools from
the extensively studied background of Brownian motion.
One may expect that this probabilistic representation
assumes a role in the investigation of Berezin-Toeplitz operators 
similar to that of the Feynman-Kac formula in the analysis of 
Schr\"odinger operators.

\subsection{Structure and Contents}

%The contents of this work split in two major parts.
%In the beginning, we study conditions for the self-adjointness
%and semiboundedness of Berezin-Toeplitz operators, which may then serve as 
%generators of strongly continuous self-adjoint semigroups.
%The second part establishes the probabilistic representation of 
%these semigroups as Wiener-regularized path integrals.
%
%The sections have the following contents: 

In Section~\ref{ch:3} we we show that a class of coherent states
is essential to the understanding of Berezin-Toeplitz quantization.
After defining Berezin-Toeplitz operators in terms
of semibounded quadratic forms, we give an abstract condition for 
their self-adjointness.
Section~\ref{ch:4} establishes a relationship between Berezin-Toeplitz
and Schr\"odin\-ger operators, which makes standard techniques
from the context of differential operators available 
to formulate more concrete conditions ensuring the
self-adjointness of a Berezin-Toeplitz operator.
The main topic of Section~\ref{ch:5} is the probabilistic representation
of semigroups generated by self-adjoint, semibounded Berezin-Toeplitz 
operators. This result is called the Daubechies-Klauder formula.
It is derived from a version of the Feynman-Kac formula for Schr\"o\-din\-ger
operators on Riemannian manifolds.
Finally, we summarize the results in Section~\ref{ch:7}
and conclude with an outlook on further developments.

% --------------------------------------------------------------------
\section{Berezin-Toeplitz Quantization from a Coherent-State
Perspective}
\label{ch:3}
% --------------------------------------------------------------------

This section explains the construction of self-adjoint operators 
according to a quantization scheme in the spirit of Berezin 
\cite{Ber72a,Ber74}.
In a geometric formulation of this scheme 
\cite{BMS94,CGR90,CGR93,CGR94,CGR95}, 
the underlying Hilbert space contains square-integrable, 
holomorphic sections in a holomorphic line bundle $\L$ with a compatible 
connection $\nabla$ over the classical phase space $\M$. 
The correspondence between the geometry of the line bundle and the classical 
phase-space structure is implicit in the fundamental assumption
that the symplectic form on $\M$ can be reconstructed as a multiple 
of the curvature associated with the connection.
%In analogy with the usual interpretation of
%probability amplitudes, the probability that a measurement will find a 
%quantum system described by such a section in a given subset of phase 
%space emerges according to the following procedure:
%First, the length of a section at any base point must be defined.
%Up to an overall constant, this is determined by asking 
%the horizontal transport determined by the connection
%to be length-preserving. Integrating the square of 
%this length against Liouville's volume form over the phase-space subset 
%in question then gives the desired probability. Hereby, the constant
%is chosen in order to normalize the resulting probability measure.

The first part of this section describes how a family of coherent 
states arises naturally with Berezin-Toeplitz quantization.
Conversely, it is possible to recover some of the additional structures 
that are imposed on the classical phase space from the presence of 
such coherent states. The details are explained
in the following exposition.  

Klauder's concept of a continuous representation 
\cite{Kla63a,Kla63b,Kla64,KM65,KMK65,MK64} 
is based on the existence of a family of orthogonal projectors 
$\{\Pi_x\}_{x \in \M}$
onto one-di\-men\-sio\-nal subspaces of a separable complex Hilbert 
space $\H$, indexed by points in a topological manifold such that
$x \mapsto \Pi_x$ is weakly continuous. If there is
a measure $\gm$ on $\M$ such that the integral 
$\int_\M \Pi_x d\gm(x) = \id_\H$
provides a weakly convergent resolution of the identity mapping $\id_\H$,
then we call each one-dimensional subspace $e(x):=\Pi_x\H$ a coherent state.
Thus, one can think of the manifold $\M$ as being embedded
in the projective Hilbert space $P\H$, the set of all one-dimensional 
subspaces of $\H$. By definition, the image of the embedding  
constitutes the family of coherent states. The identification
of collinear vectors in $\H$ to describe a (pure) quantum state
induces additional structures on $\M$. 

Since $P\H$ is the base manifold of a bundle $P: \H\setminus\{0\} \to P\H$, 
where the projection $P$ maps any nonzero vector in $\H$ to the
one-dimensional subspace it generates, the embedding
of $\M$ pulls back the fibers $\pi^{-1}(\{x\}) := P^{-1}(\{e(x)\}), x \in \M$.
To make $\M$ the base manifold of a complex line bundle,
the missing zero vector must be inserted in every fiber
$\L_x := \pi^{-1}(\{x\}) \cup \{0\}$ and thus a bundle is created 
with total space $\L=\bigcup_{x \in \M} \L_x$ and projection $\pi$. 
If we suppose that the linear hull of $\L$ is dense in $\H$,
then the linear functional $\vartheta_v: \psi \mapsto (v,\psi)$ 
restricted to $\psi \in \L$ provides a representation of $v \in \H$ 
as a function on $\L$ that is complex linear in the fibers. 
If $\M$ is a differentiable manifold and the mapping $x \mapsto \Pi_x$
is in some sense smooth, then as subsets of $\H$, the fibers in the total 
space $\L$ inherit additional
features. 
%For example, the scalar multiplication $\psi \mapsto c \psi$
%of vectors $\psi \in \H$ provides a natural fiber-preserving group action of 
%nonzero complex numbers $c \in \C^\times$.
The scalar product $(\centerdot,\centerdot)$ serves simultaneously as a 
Hermitian metric on both the total space $\L$ and the tangent space $T\L$. 
%By definition, these metrics are invariant under all endomorphisms 
%mapping $\L$ to $\L$ that are restrictions of unitary transformations 
%on $\H$, such as the scalar multiplication of all vectors
%by a unimodular number $c, \abs{c}=1$. 
The notion of horizontal transport passes from $\H$ to $\L$, 
which takes a smooth curve $\zeta: \R \to \M$ together with
a starting point $\hat \zeta(0)$ in $\pi^{-1}(\zeta(0))$ 
and produces the lifted curve  
$\hat \zeta$ in $\L$ by moving in an infinitesimal time step $dt$
from $\hat \zeta(t), t \in \R$, to the orthogonal projection of 
$\hat \zeta(t)$ onto the space $e(\zeta(t+dt))$. 
In fact, this way the norm of a horizontally transported
vector in the fiber is left invariant while its base point moves 
along the curve in $\M$. In other words, the connection on the bundle 
corresponding to the horizontal transport is compatible with the 
Hermitian structure.

Berezin-Toeplitz quantization realizes a class of
such continuous representations in a setting that is
familiar in algebraic geometry \cite{GH78}: If $\L$ is
a holomorphic line bundle over a K\"ahler manifold,
then the curvature of the line bundle is a closed 
two-form \cite{Zha00}. This two-form is up to an imaginary factor 
assumed to be equal to the symplectic form on $\M$.
The Hilbert space chosen by Berezin-Toeplitz 
quantization is the space of holomorphic sections that are 
square-integrable with respect to Liouville's measure, the 
maximal exterior power of the curvature form.
These requirements are needed to show a correspondence principle
for compact $\M$ \cite{BMS94,Sch98}. Unfortunately, they also restrict the 
universality of Berezin-Toeplitz quantization. Not all symplectic 
manifolds can be equipped with a compatible complex structure, 
and even less may be obtained as the base manifold of a holomorphic 
line bundle such that its curvature is a constant multiple of 
the original symplectic form \cite{Sch98}. 
%Because of the demonstrated invariance properties of the Hermitian metric
%on $T\L$, the tangent bundle $T\M$ inherits a Fubini-Study type metric
%\cite[Appendix 3]{Arn89}. It is straightforward to check that the 
%imaginary, skew-symmetric part of the Fubini-Study metric is closed
%and therefore constitutes yet another way to derive a symplectic form
%from the embedding, which makes $\M$ a K\"ahler manifold. 
%It turns out that this symplectic form may or may not 
%coincide up to a constant factor with the curvature of the 
%line bundle \cite{CGR93}. 
In order to provide a resolution of the identity $\id_{\H}$ according to
$\int_\M \Pi_x d \gm(x) = \id_{\H}$, the measure $\gm$ is chosen
as a locally rescaled version of the Liouville form, see \cite{CGR93}.
More generally, Berezin-Toeplitz quantization
maps the classical observable represented by a 
bounded real-valued function
$f:$ $\M \to \R$ to the self-adjoint operator obtained from
$T_f :=$ $\int_\M f(x) \Pi_x d\gm (x)$.
In both cases, the integral converges in the strong sense.
The quantization of dynamics is then realized 
with the unitary group $\{e^{-itT_f}\}_{t\in \R}$
that results from choosing $f$ as the generator of classical 
time evolution.

\subsection{Hilbert Spaces of Square-Integrable, Holomorphic Sections}%

\begin{defn} 
  Let us assume that a complex line bundle $\L \stackrel{\pi}{\to} \M$ is 
equipped with a Hermitian metric $h=\{h_x\}_{x \in \M}$ on its fibers.
To be precise, for each base point $x \in \M$ there is a
sesquilinear metric $h_x: \L_x\times\L_x \to \mathbb C$ on the associated 
fiber $\L_x$. By  convention, each $h_x$ is conjugate linear in the first 
argument. We will only consider finite-dimensional manifolds, 
$n := \mathrm{dim}_\C \M < \infty$.
Given a measure $\meas$ on $\M$ 
we may define an inner product
\begin{equation}
  \label{eq:IP}
  (\psi,\phi) := \int_\M h(\psi,\phi)\, d\meas 
\end{equation}
for sufficiently regular sections $\psi$ and $\phi$,
where $h(\psi,\phi)$ is interpreted as the function
$x \mapsto h_x(\psi(x),\phi(x))$.
\end{defn}

\begin{rem}
  In the definition of the inner product, $h$
and $\meas$ can be combined to a Hermitian-metric
valued measure, hereafter denoted by $h\meas$. Indeed, this is a more 
appropriate way to view the definition, since the redundancy of 
rescaling $h$ while changing $\meas$ to compensate accordingly
is manifest in the notation. 
\end{rem}

\begin{defn}
The linear space of sections in $\L$ will be denoted as $\Gamma_\L(\M)$.
The subspace of square-integrable sections on a complex line bundle 
$\L$ over a base manifold $\M$ is denoted by
\begin{equation}
  \label{eq:L2space}
  L^2(h \meas) := \Bigl\{ \psi \in \Gamma_\cL(\M): 
                         \int_\M h(\psi,\psi)\, d\meas < \infty \Bigr\} \, .
\end{equation}
When $\L$ is a holomorphic line bundle, we define
the generalized Bergman space $L^2_{hol}(h \meas)$ as
the space of all holomorphic sections in $L^2(h \meas)$.
\end{defn}

\begin{remarks}
Equipped with the previously defined inner product, 
the space $L^2(h \meas)$ containing all square-integrable sections 
becomes a Hilbert space in the usual way by identifying
sections that differ up to sets of $h\meas$-measure zero.

If $\L$ is a holomorphic line bundle and 
$\meas$, interpreted as a volume form, and
$h$ are everywhere non-degenerate 
and smooth, then the generalized Bergman space 
$\Bg$ is a space of functions that
may be identified with a Hilbert-subspace of $L^2(h \meas)$.
An outline of the completeness proof is given in 
Appendix~\ref{app:A}.

  It may happen that $L^2_{hol}(h \meas)$ only contains the zero section.
Therefore, results about the dimensionality of this space 
are of  fundamental interest. For the case of compact $\M$, 
see \cite{BMS94}.
\end{remarks}

\begin{lem}
Given a vector $u$ in a fiber above $x:=\pi(u)$, the 
point evaluation 
\begin{equation}
\begin{split}
  \vartheta_u: L^2_{\hol}(h \meas) &\longrightarrow \C \\
                          \psi &\longmapsto h_{x}(u,\psi(x))   
  \label{eq:pevfnal}
\end{split}
\end{equation}
defines a bounded linear functional, and by the Riesz
representation theorem this evaluation can be realized as
an inner product $\tilde \psi(u):=(e_u, \psi) = \vartheta_u(\psi)$
with a section $e_u \in \Bg$. Two such sections form a 
kernel function $k(u,v):=(e_u,e_v)$ that is defined on $\L \times \L$
and sesquilinear in the fibers.
\end{lem}
\begin{proof}
  The detail that mostly deserves explanation is the 
boundedness of $\vartheta_u$. To verify this, we
choose a local trivialization $\xi$ around the fiber generated by $u$,
mapping $\pi^{-1}(U) \subset \L$, the subset of $\L$ above an open 
set $U$ to $ V\times \C$,
with an open ball $V \subset \C^n$ having the first component of
$\xi(u)$ as the center. 

Given a convergent sequence of sections
$\{\psi^{(l)}\}_{l \in \N}$, we use as in Appendix~\ref{app:CTB} 
the mean value property of the associated holomorphic functions 
on $V$ to bound the value of $\vartheta_u(\psi^{(l)})$ 
by a constant times the $L^2$-norm of $\psi^{(l)}$. Since the sequence 
has the Cauchy property, $\vartheta_u(\psi^{(l)})$ is also Cauchy, and 
therefore convergent. 

The sesquilinearity of $k$ results from the conjugate-linear scaling 
property $e_{cu} = \ol c \, e_u$ for any $c \in \C$ and $u \in \L$. 
\end{proof}

\begin{defn}
A Schwartz kernel in a complex line bundle $\L$ is a family
of linear mappings $\{\mathcal{S}(x,y):$ $\L_y \to \L_x\}_{x,y \in \M}$, that
is, $\mathcal{S}(x,y)$ is linear in vectors with base point $y$ and has 
as its values vectors at $x$.   
If $\mathcal{S}(x,y)$ is jointly continuous in $x$ and $y$, then
it can be interpreted as continuous section  in the bundle 
$\L \otimes \L^* \to \M \otimes \M$,
where $\L^*$ is the dual bundle associating with each $x \in \M$
the space of complex linear forms on $\L_x$.
\end{defn}

%% recall \meas!!
%% CHANGE
\begin{prop}
  The Schwartz kernel $K$ given
in the terminology of the preceding lemma by $K(x,y) v = e_v(x)$ 
for $v \in \pi^{-1}(y)$ is jointly continuous in $x$ and $y$.
We will call $K$ the reproducing kernel of $L^2_{\hol}(h \meas)$
because all $\psi \in L^2_{\hol}(h \meas)$ satisfy the identity
\begin{equation}
  \label{eq:repkernel}
  \psi(x) = \int_\M K(x,y) \psi(y) d\meas(y) \, .
\end{equation}
\end{prop}
\begin{proof}
The joint continuity follows from
the continuity of $e_v$ in $v$ and the 
uniform convergence of Cauchy sequences
in $L^2_{\hol}(h \meas)$. These properties may
be obtained using the definition of $e_v$ 
via \eq{pevfnal} and the
argument in Appendix~\ref{app:CTB}.

To derive \eq{repkernel}, we consider in a first step the adjoint map
$(K(x,y))^*: \pi^{-1}(x) \to \pi^{-1}(y)$, in the usual way defined
by $u \mapsto h_x(K(x,y) v,u)v$, independent of the choice of
a normalized vector $v \in \pi^{-1}(y), \norm{v}=1$.
We claim that $(K(x,y))^*=K(y,x)$, which means for all $u,v$
in fibers above $x$ and $y$, respectively, the equation
$h_x(u, K(x,y)v) = h_y(K(y,x)u,v)$ holds.
To simplify the following calculation, we assume that $u$ and $v$ 
are normalized; the general case follows by rescaling.
\begin{align}
  \label{eq:Kadjoint}
  h_x(u,K(x,y)v) &= h_x(u, e_v(x)) = h_x(u,\tilde e_v(u) u) \\
  &= \tilde e_v(u) = \ol{ \tilde e_u(v) } \\
  &= h_y(\tilde e_u(v) v, v)  = h_y(e_u(y),v) \\
  &= h_y(K(y,x)u,v)
\end{align}
The second step for the derivation of \eq{repkernel}
uses again a normalized vector $u$ above $x$,
\begin{align}
  \label{eq:Reppty}
  \psi(x)&= \tilde \psi(u) u = (e_u,\psi)\,u
   = \int_\M h_y(e_u(y),\psi(y)) \,u \, d\meas(y) \\
  &= \int_\M h_y(K(y,x)u,\psi(y)) \,u\, d \meas(y) \\
  &= \int_\M h_x(u,K(x,y)\psi(y)) \,u \, d \meas(y) \\
  &= \int_\M K(x,y)\psi(y) \, d \meas(y) \, .
\end{align}
\end{proof}

\begin{itcomment}
  One of the goals in this work is to find a formula for this kernel.
In principle, one could follow a Gram-Schmidt orthogonalization procedure,
construct an orthonormal basis of sections $\{\eta_l\}_{l \in \N}$
and then express the reproducing kernel as a series
$K(x,y)=\sum_l \eta_l(x) h_y(\eta_l(y),\centerdot)$ that terminates
after finite terms or converges
uniformly on compact sets in $\M \times \M$. However, this procedure is too abstract to
show how the geometry of $\L$ shapes the kernel.
We will therefore present an alternative strategy, expressing
$K$ in a probabilistic way.
\end{itcomment}

\begin{consequence} \label{con:intkernels}
  If $h\meas$ is smooth and nowhere degenerate,
then any bounded operator $B$ on $L^2_{\hol}(h\meas)$ possesses a
sesqui-analytic integral kernel
$B(x,y)$ that is characterized by the equation 
$h_x(u,B(x,y) v)= (K(\centerdot,x)u,$ $B K(\centerdot,y)v)$, 
and the image of $\psi \in L^2_{\hol}(h\meas)$ is expressed as
\begin{equation}
  \label{eq:Bpsi}
  B \psi(x) = \int_\M   B(x,y) \psi(y) \, d\meas(y) \, .
\end{equation}
\end{consequence}
\begin{proof}
  That $B(x,y)$ is indeed an integral kernel results from the 
reproducing property \eq{Reppty} and Fubini's theorem. The sesqui-analyticity
of $B(x,y)$ follows because the mapping 
$v \mapsto K(\centerdot, \pi(v))v=e_v$ into $L^2_{\hol}(h\meas)$
is antiholomorphic.
\end{proof}

\begin{rem} 
Since the right-hand side of equation \eq{Bpsi} 
is defined even for $\psi \in L^2(h\meas)$, any bounded 
operator extends naturally 
via its integral kernel to all of $L^2(h\meas)$. From this point of 
view, $K(x,y)$ is the integral kernel 
of an orthogonal projection operator, henceforth also called $K$, that maps 
$L^2(h\meas)$ onto $L^2_{\hol}(h\meas)$.
\end{rem}

\subsection{Berezin-Toeplitz Operators Defined via Quadratic Forms}

In the remaining text, we assume that $h$ and $\meas$ are smooth and
non-degenerate to ensure that $\Bg$ is complete.

\begin{defn}
 Given the Hilbert space $\Bg$ and a real-valued function $f: \M \to \R$, 
we consider the sesquilinear form
\begin{align}
  \label{eq:Qform}
  \t_f: \Q( \t_f) \times \Q(\t_f)  &\longrightarrow \C \\
       (\psi, \phi) &\longmapsto \int_\M f(x) h_x(\psi(x), \phi(x)) d\meas(x) 
   \label{eq:tfdef}
\end{align}
with form domain 
\begin{equation}
  \label{eq:tfdom}
  \Q(\t_f) := \Bigl\{ \psi \in \Bg: \int_\M \abs{f(x)} h_x(\psi(x),\psi(x)) d\meas(x) 
                                 < \infty \Bigr\} \, .
\end{equation}
When referring to $\t_f$ as a quadratic form, it is really the 
function $\psi \mapsto \t_f(\psi,\psi)$ that is meant.
\end{defn}

\begin{defn}
  Given a real-valued, bounded function $f: \M \to \R$, the 
 form $\t_f$ specified in the preceding definition is
bounded and symmetric. Therefore, it is  
associated with a self-adjoint operator $T_f$ satisfying $(\psi,T_f\psi)=\t_f(\psi,\psi)$
for all $\psi \in \Bg$. In the context of generalized Bergman spaces, we 
call $T_f$ a self-adjoint Berezin-Toeplitz operator and 
the function $f$ its symbol.
\end{defn}

%%% CHANGED 3/5/01
\begin{itremarks}
The original definition according to Berezin \cite{Ber72a,Ber74} 
and its geometric interpretation by Cahen, Gutt, Rawnsley and others 
\cite{BMS94,CGR90,CGR93,CGR94,CGR95}
do not refer to sesquilinear forms. Indeed, for bounded symbols
the approach chosen here offers no new insights.

However, the use of sesquilinear forms is convenient for 
the construction of semibounded Berezin-Toeplitz operators
described in the remaining part of this section.
The implicit goal is to find a large class of possibly unbounded 
symbols $f$ that lead to closed, semibounded quadratic forms $\t_f$ 
and thus yield unique self-adjoint Berezin-Toeplitz operators $T_f$ 
via the Friedrichs construction characterized by equation \eq{FRT}.
In fact, this goal leads the discussion 
from abstract conditions ensuring the semiboundedness of $T_f$
to a more concrete class of admissible symbols presented in the 
next section.
\end{itremarks}
%%% CHANGED 3/5/01

\begin{lem} \label{thm:KLMN}
  If the form $\t_{f^+}$ belonging to the positive part $f^+:$ $x \mapsto$
$\max\{f(x),$ $0\}$ of a  
function $f: \M \to \R$ is
densely defined and the negative part 
$f^-: x \mapsto \max\{-f(x),0\}$
can be incorporated in $\t_f$
as a form-bounded perturbation, meaning
\begin{equation}
  \label{eq:fbd}
  \t_{f^-}(\psi,\psi) \leq c_1 \, \t_{f^+}(\psi,\psi) + c_2 \norm{\psi}^2 
\end{equation}
with a relative form bound $c_1<1$ and a constant $c_2 \ge 0$,
then $\t_f$ is closed on $\Q(\t_f) =\Q(\t_{f^+})$
and has a lower bound $c \in \R$,
such that $\t_f(\psi,\psi) \ge c \norm\psi^2$. 
\end{lem}
\begin{proof}
The first part of the proof is to show that
the sesquilinear form belonging to a non-negative function $f \ge 0$  
is closed, in other words,
we need to show that $\Q(\t_f)$, equipped with the form-norm 
$\norm{\bullet}_{\t_f}$ defined by
  \begin{align}
    \label{eq:tnorm}
       \norm{\psi}_{\t_f} := ( \t_f(\psi,\psi) + \norm\psi^2 )^{1/2} 
           & \text{\ \ for\ \ } \psi \in \Q(\t_f) \, ,
  \end{align}
is complete.

Suppose $(\psi_l)_{l \in \mathbb N}$ is a Cauchy sequence with respect to 
the form-norm. 
Due to the estimate $\norm\psi \leq \norm{\psi}_{\t_f}$
the sequence is convergent in $\Bg$, $\psi_l \to \psi$. Using pointwise 
convergence and Fatou's lemma, we have
$
   \norm{\psi - \psi_l}_{\t_f} \leq  $ 
$\liminf_{k \to \infty} 
\norm{\psi_k - \psi_l}_{\t_f}
$
and therefore the sequence $(\psi_l)_{l \in \mathbb N}$ converges with 
respect to the form-norm.

The remaining part of the proof is the so-called KLMN theorem, 
see \cite{Sim71} or \cite[Theorem X.17]{RS75}.
It goes back to works of Kato \cite{Kat55}, Lax and Milgram \cite{LM54},
Lions \cite{Lio61} and Nelson \cite{Nel64b}.
\end{proof}

\begin{prop} \label{fact:FRT}
If the form $\t_f$ is closed and has the greatest lower bound $c \in \R$, 
then it belongs to 
a unique self-adjoint operator $T_f$ that is characterized in terms of the 
square-root $\sqrt{T_f - c}$ satisfying
\begin{equation} \label{eq:FRT}
   (\sqrt{T_f - c}\, \phi,\sqrt{T_f - c}\, \psi) + c(\phi,\psi) 
       = \t_f(\phi,\psi) 
\end{equation}
for all $\phi$ and $\psi$ in the domain $\D(\sqrt{T_f - c}) = \Q(\t_f)$.
\end{prop}
\begin{proof}
Again, we refer to the literature \cite[Theorem VIII.15]{RS80} or 
\cite[Theorem 5.36]{Wei80} 
for the proof of this result which we call the Friedrichs construction.
\end{proof}

\begin{remarks}
As a special case of Consequence~\ref{con:intkernels}, 
when $f$ is a bounded function,
 $T_f$ has an integral kernel $T_f(x,y)$
characterized by $h_x(u,T_f(x,y)v)= (K(\centerdot, x)u,$ $f K(\centerdot, y)v)$, where 
$u,v \in \L$ have base points $x$ and $y$, and
the scalar product is taken in $L^2(h\meas)$.  

For $\psi \in \D_{\mbox{\scriptsize\it min}}(T_f):= 
                   \{ \psi \in \Bg, f \psi \in L^2(h\meas) \}$, the identity
$T_f \psi = K (f\psi)$ relates $T_f$ to the traditional way 
of defining a Berezin-Toeplitz operator as a composition of
a multiplication operator with the orthogonal 
projection $K$. However, it may happen that $\D_{\mbox{\scriptsize\it min}}(T_f)$
does not include all of $\Bg$, although the operator $T_f$ is bounded.

A disadvantage of defining $T_f$ by a semibounded 
form is that in general, nothing is known about its domain. 
The situation is different if a
domain of essential self-adjointness can be identified for $T_f$. 
Such situations have been investigated in detail 
\cite{Cic96,JS94} for the case of the so-called Fock-Bargmann 
space.

The definition of Berezin-Toeplitz operators clearly does 
not rely on the validity of a correspondence principle,
and we will also not need to refer to it hereafter. 
Because of its physical importance, we mention that
in the special setting of holomorphic line bundles over 
homogeneous or compact K\"ahler manifolds,
the Berezin-Toeplitz operators defined on
the Hilbert space $L_\hol^2(h\meas)$, with $\meas$ being the
Liouville form associated with the bundle curvature, are known to
observe a correspondence principle, see \cite{Ber74,Per86}
or \cite{BMS94,Sch98}. Moreover, in the compact case 
the same kind of classical asymptotics can be proved
for more general almost-complex manifolds \cite{BU96}.
\end{remarks}

% ---------------------------------------------------------------------
\section{Self-Adjoint Berezin-Toeplitz Operators as Monotone Limits of 
         Semibounded Schr\"odinger Operators}
\label{ch:4}
%\markboth{\thechapter. Self-Adjoint Berezin-Toeplitz Operators}{}
% ---------------------------------------------------------------------

%% Riemannian Volume element = bundle curv^n!?

The main motivation for this section is to relate Berezin-Toeplitz 
and Schr\"o\-din\-ger operators. 
An important application concerns the transfer of well-known 
self-adjointness criteria to the setting of Berezin-Toeplitz operators. 
In this section and the following one we derive conditions that are 
more accessible than verifying the abstract form-boundedness of 
$T_{f^-}$ with respect to $T_{f^+}$ according to in\-equa\-li\-ty~\eq{fbd}.

At first, the Riemannian structure seems to be an auxiliary element 
that is not needed in the definition of Berezin-Toeplitz operators 
according to the prescription of the preceding section.
However, continuing the line of thought in
the introductory remarks given there, we note that
due to the coherent-state embedding $x \mapsto e(x):=\{e_u: u \in \L_x\}$
the inner product of the Hilbert 
space provides a natural metric on $T\L$, which by its invariance
under scalar multiplication in the fibers passes as
Fubini-Study type metric \cite[Appendix 3]{Arn89} to the
tangent bundle $T\M$. It is straightforward to check that the 
imaginary, skew-symmetric part of the Fubini-Study metric is closed,
which makes $\M$ a K\"ahler manifold. The real part of this metric 
can then be used to define a Riemannian structure. In short,
a Riemannian metric is present as a consequence of the
quantization prescription.   

In the following, we consider Hilbert spaces $L^2_\hol(hm)$
of square-integrable holomorphic sections in a holomorphic Hermitian line
bundle $\L$ that has a base manifold $\M$ with a K\"ahler metric.
A priori, the natural volume measure $m$ associated with the real 
part of the K\"ahler metric need not be in a prequantum 
relation~\eq{pqrel} with the curvature of $\L$.

\subsection{Bochner's Laplacian and its Relation to the Holomorphic Laplacian}

Several Laplacians will be introduced in this section, 
each one is characterized by an associated 
positive definite quadratic form. Later, Schr\"odinger
operators will arise from perturbations of these forms.

\begin{conventions}
By default, 
$\M$ is always a $d$-dimensional Riemannian manifold, and
whenever it appears in conjunction with the holomorphic 
line bundle $\L$, it is tacitly understood to be the 
base manifold, with $d=2n$. 
The Hermitian metric $h$
on $\L$ and the natural volume measure $m$ on $\M$ are
as forms assumed to be smooth and non-degenerate.
The class of smooth vector fields that are at each point 
real-valued differential operators on real-valued, smooth 
functions is written as $\Upsilon_\R(\M)$. The complexified 
version is written
as $\Upsilon(\M)$. Whenever a smooth vector field $Y$ 
given on an open set $U \subset \M$ vanishes on
all antiholomorphic functions in $U$, we write $Y \in \Upsilon^{(1,0)}(U)$,
and if this happens for all holomorphic ones $Y \in \Upsilon^{(0,1)}(U)$.  
We will not distinguish between a Riemannian metric $g$
on the tangent bundle $T\M$ and its sesquilinear
extension to the complexified tangent bundle $T^{\mkern2mu\C}\M$,
as usual conjugate linear in the first argument. Similarly, the 
Levi-Civita connection $\Cov$ is made  complex linear
on $T^{\mkern2mu\C}\M \times T^{\mkern2mu\C}\M$, and the divergence 
$\mathrm{div}$ is thus defined as the trace $\mathrm{Tr}^{T\M}b_Y$
for all $Y \in \Upsilon(\M)$ with the sesquilinear form 
given by $b_Y: (X, Z) \mapsto (X, \Cov_Z Y)$. The gradient
of a function $f$ is a vector field denoted as $\mathrm{grad} f$.

With a view to Lemma~\ref{lem:cutoff}, from now on all 
manifolds are  tacitily assumed to be pathwise connected. 
\end{conventions}

\begin{defn}
The operator obtained by the Friedrichs construction
corresponding to the closure of the quadratic form 
\begin{equation}
  \label{eq:Laplace}
   \cE(f,f)
  := \int_\M g(\mathop{\mathrm{grad}} f, \mathop{\mathrm{grad}} f) dm
\end{equation}
with initial form domain $C_c^\infty(\M)$ is called the
negative Dirichlet Laplacian $-\Delta$ on $L^2(m)$.

Suppose $\L$ is a Hermitian line bundle with a compatible
connection $\nabla$.
The negative Bochner Laplacian $-\Delta^\cL$ on $L^2(hm)$ arises via
the Friedrichs construction from  
\begin{equation}
  \label{eq:BochnerLaplacian}
  \cE^\cL(\psi,\psi) := \int_\M \mathop{\mathrm{Tr}^{T\M}}h(\nabla \psi, \nabla \psi) dm 
\end{equation}
defined on $\Cinfty_{c\L}(\M)$, the space of smooth sections with 
compact support. Hereby, the trace operation is defined as before by 
choosing an orthonormal basis 
$\{E_k\}_{k=1}^d$ in each $T_x\M$ such that 
$\mathop{\mathrm{Tr}^{T\M}}h(\nabla \psi, \nabla \psi)
 = \sum_{k=1}^d h(\nabla_{E_k} \psi, \nabla_{E_k} \psi)$.
\end{defn}

\begin{lem} \label{lem:cutoff}
  Every complete Riemannian manifold $\M$ admits a localizing sequence
of smooth cut-off functions with a uniformly attenuated gradient bound.
This means, there is an increasing sequence $\{\eta_l\}_{l \in \N}$
of smooth functions $\eta_l$ pointwise converging to unity, $\eta_l(x) \nearrow 1$
for all $x \in \M$, each $\eta_l$ has compact support, and the 
uniform gradient bound $g(\grad \eta_l, \grad \eta_l) \leq C_l$ holds for some sequence
$\{C_l\}_{l \in \N}$ of positive numbers $C_l\ge 0$ converging to zero.
\end{lem}
\begin{proof}
  The construction uses a result by Greene and Wu \cite[Corollary to Proposition 2.1]{GW79},
by which one may approximate the distance from a fixed point $y \in \M$
with a smooth function. To be precise, one obtains
a smooth function $\upsilon: \M \to \R$ such that $\norm{\grad \upsilon} < 1$
and $\abs{\upsilon(x) - \mathrm{dist}(x,y)}<1$ for all $x \in \M$.

For the construction of the cut-off functions, we pick a real-valued
smooth function $\eta: \R \to [0,1]$ that is bounded above and below
by characteristic functions $\chi_{[-1,1]} \leq \eta \leq \chi_{[-2,2]}$,
ensuring compact support in the interval $[-2,2]$.
The composition $\eta_l(x):=\eta(\frac 1 {2^l} \upsilon(x))$
then defines an increasing sequence of smooth functions $\eta_l\nearrow 1$ with 
the gradient bound
\begin{equation}
  \label{eq:gradbd}
  \grad \eta_l = \frac 1 {2^l} \eta'(\frac 1 {2^l} \upsilon (x) )\grad \upsilon (x)
               \leq \frac 1 {2^l} \max_{r \in \R} \abs{\eta'(r)} \, .
\end{equation}
In addition, due to the completeness of the manifold, the support of each $\eta_l$ 
is compact since it is contained in the closed set $\upsilon^{-1}([-2^l,2^l])$.  
\end{proof}

\begin{thm} \label{thm:esssa}
  If the Riemannian manifold $\M$ is complete, then $-\Delta$ is
essentially self-adjoint on $C_c^\infty(\M)$.
The same holds for $-\Delta^\cL$ in 
a Hermitian line bundle $\L$ with a compatible connection $\nabla$.
Moreover, $-\Delta^\cL$ has
$\Cinfty_{c\L}(\M)$ as a domain of 
essential self-adjointness.  
\end{thm}
\begin{proof}
  It is sufficient to show this for 
$\Delta^\cL$, since $\Delta$ can be considered as the Bochner Laplacian
on the trivial bundle $\M \times \C$ with the obvious Hermitian structure.
We adapt Davies' treatment of the Dirichlet Laplacian 
\cite[Theorem 5.2.3]{Dav89} in combination with
the localizing sequence of cut-off functions described in the 
preceding construction.

The essential self-adjointness of $-\Delta^\cL$ is by its positivity 
equivalent \cite[Theorem X.26]{RS75}
to having only the zero vector in
the orthogonal complement of $(- \Delta^\cL + 1)\Cinfty_{c\L}(\M)$.

Suppose there is a nonzero vector $u \bot (- \Delta^\cL + 1)\Cinfty_{c\L}(\M)$,
in other words the equation $\Delta^\cL u = u$ has a weak solution $u \in L^2(hm)$.
Using the localizing sequence $\{\eta_l\}_{l \in \N}$ described above,
we may estimate 
% smoothness of u needed here!!
\begin{align}
  \label{eq:essadj}
  0 &\ge - \norm{ \eta_l u}_2^2 = \cE^\cL(\eta_l^2u,u)
    = \int_\M \sum_{k=1}^d h(\nabla_k \eta_l^2 u, \nabla_k u) dm \\
    &= \int_\M \sum_{k=1}^d 2 \eta_l E_k(\eta_l) h(u, \nabla_k u) dm \nonumber 
              + \int_\M \sum_{k=1}^d \eta_l^2 h(\nabla_k u, \nabla_k u) dm \, .
\end{align}
The last term is positive and we conclude that it must be bounded by
\begin{align}
  \label{eq:essadjII}
  \int_\M \sum_{k=1}^d \eta_l^2  h(\nabla_k u, \nabla_k u) dm 
   &\leq 2 \int_\M \sum_{k=1}^d \eta_l \abs{E_k(\eta_l)} \abs{h(u, \nabla_k u)} dm \\
  &\leq 2 \int_\M \sum_{k=1}^d \eta_l \abs{E_k(\eta_l)} \sqrt{h(u,u) 
            h(\nabla_k u, \nabla_k u)} \, dm\\
  &\leq 2 \int_\M \eta_l \norm{\grad{\eta_l}}_\infty \sqrt{h(u,u) 
             {\textstyle\sum_k} h(\nabla_k u, \nabla_k u)} \, dm \, \quad
\end{align}
where the Cauchy-Schwarz inequality has been used repeatedly.
With the abbreviation $c_l:= \eta_l \sqrt{{\textstyle\sum_k}h(\nabla_k u, \nabla_k u)}$,
we obtain
\begin{equation}
  \label{eq:cnsqd}
  \int_\M c_l^2 dm \leq 2 \int_\M c_l 
          \norm{\grad \eta_l}_\infty \sqrt{h(u,u)} \, ,
\end{equation}
and after using the Cauchy-Schwarz inequality again,
\begin{equation}
  \label{eq:CSagain}
  \norm{c_l}_2^2 \leq 2 \norm{c_l}_2 \norm{\grad \eta_l}_\infty \norm{u}_2 \, .
\end{equation}
To avoid confusion, $\norm{\centerdot}_2$ denotes the $L^2$-norm and the term
$\norm{\grad \eta_l}_\infty$
the essential supremum of the Riemannian length of $\grad \eta_l(x)$ over $x \in \M$.
This last inequality involves finite quantities on both sides, because
$u$ is a smooth function by an argument related to Sobolev norms as in 
Appendix~\ref{app:hk}.
The properties of the localizing sequence $\{\eta_l\}$ imply
that the right-hand side
approaches zero in the limit $l \to \infty$. Therefore, by
Fatou's lemma $\cE^\cL(u,u)=0$ or $\nabla u = 0$, which is in contradiction 
to the assumption  $\Delta^\cL u = u \neq 0$.
\end{proof}

Now we investigate the interplay between
Riemannian and complex structures on $\M$.
The fundamental ingredient is the assumption that
the connection $\nabla$ is compatible with the 
Hermitian metric and the holomorphic structure,
that is, $\nabla_X \psi = 0$ for all
locally antiholomorphic vector fields $X$ 
and locally holomorphic sections $\psi$.

\begin{defn} \label{def:Zj}
Suppose we pick a local section in the orthonormal frame bundle
of $T^{(0,1)}\M$, which 
means in a sufficiently small open set $U\subset \M$, we have 
antiholomorphic vector fields 
$\ol Z_1, \ol Z_2,$ $\dots$ $\ol Z_{d/2} \in \Upsilon^{(0,1)}(U)$ 
that are orthonormal, $g(\ol Z_k,\ol Z_l)=\delta_{kl}$.
For a section $\psi$, the value of the antiholomorphic trace
$\mathop{\mathrm{Tr}^{(0,1)}} h(\nabla \psi, \nabla \psi) 
  :=  \sum_{k} h(\nabla_{\ol Z_k} \psi, \nabla_{\ol Z_k} \psi)$ 
depends on the metric and the connection $\nabla$, not on the particular choice 
of orthonormal antiholomorphic vector fields.
Therefore, we may define the negative holomorphic Laplacian $-\Delta^{(0,\bullet)}$, 
in a manner analogous to the previous definitions 
as the operator corresponding to the closure of the quadratic form 
\begin{equation}
  \label{eq:Delta_alpha}
  \cE^{(0,\bullet)}(\psi,\psi) := \int_\M \mathop{\mathrm{Tr}^{(0,1)}} 
                                      h(\nabla \psi, \nabla \psi) dm
\end{equation}
initially defined on sections $\psi$ in the domain $\Cinfty_{c\L}(\M)$.
\end{defn}

\begin{rem}
  Let $g$ be a Riemannian metric on a complex manifold $\M$ and $\Cov$ its 
Levi-Civita
connection. It is straightforward to check in local coordinates 
\cite[Proposition 7.14]{Zha00} that $g$ is the real part of a K\"ahler metric
if and only if it is compatible
with the almost complex structure $J$ and if $\Cov$ preserves the splitting
of $\Upsilon(\M)$ into holomorphic and antiholomorphic parts, that is,
$\Cov_X JY = J \Cov_X Y$ for all $X,Y \in \Upsilon(\M)$.
\end{rem}

\begin{prop} \label{prop:WBF}
Let $g$ be the real part of a K\"ahler metric on the
the $d$-dimensional base manifold $\M$ of a holomorphic line bundle $\L$,
and assume the Bochner and holomorphic Laplacians are defined as above.
Then a Weit\-zen\-b\"ock-type formula relates both Laplacians
\begin{equation}
  \label{eq:holomorphic2Bochner}
  \Delta^{(0,\bullet)}= \frac 1 2 \bigl(\Delta^\cL - \rho  \bigr)\, 
\end{equation}
with a zeroth-order term $\rho$.
Given an antiholomorphic orthonormal frame $\{\ol Z_k\}_{k=1}^{d/2}$ 
of  $T^{(0,1)}\M$,
the term $\rho$ is expressed as
$\rho(x) \psi(x) = \sum_{k=1}^{d/2} R_{\ol Z_k,Z_k} \psi(x)$. 
\end{prop}
\begin{proof}
The first step of the proof is to identify 
$-\Delta^{(0,\bullet)}$ and $\Delta^\L$ as differential operators
when acting on a smooth, compactly supported section 
$\psi \in C^\infty_{c\L}(\M)$. According to the 
usual derivation
\cite{BGV92} we find
\begin{equation}
  \label{eq:BochnerLapDO}
  \Delta^\cL \psi= \sum_{k=1}^d \nabla_{E_k} \nabla_{E_k} \psi 
                 - \nabla_{\Cov_{E_k} E_k} \psi \, 
\end{equation}
and
\begin{equation}
  \label{eq:holLaplacianDO}
  \Delta^{(0,\bullet)} \psi = \sum_{k=1}^{d/2}  \left( \nabla_{Z_k} \nabla_{\ol Z_k} - 
                                        \nabla_{\Cov_{Z_k}\ol Z_k} \right) \psi \, . 
\end{equation}
Here, $\{E_k, J E_k\}_{k=1}^{d/2}$ is a local orthonormal frame
in $T\M$ and the antholomorphic frame $\{\ol Z_k\}_{k=1}^{d/2}$ 
is obtained via $\ol Z_k = \frac 1{\sqrt 2}(E_k+iJE_k) \in T^{(0,1)}\M$.
The derivation relies on the compatibility of the connection
$\nabla$ with the Hermitian metric, the resolution of the identity 
$
  \sum_{k=1}^{d/2} \left( (\centerdot,E_k)E_k + (\centerdot, JE_k)JE_k \right)
   = \id_{T\M}
$
and the compatibility between the almost complex structure $J$ and
the Levi-Civita connection $\Cov$. 
Now the claimed relationship  \eq{holomorphic2Bochner} follows from
\eq{BochnerLapDO} and \eq{holLaplacianDO} with
the torsion-free property of the Levi-Civita connection $\Cov$ that
implies $[\ol Z_k, Z_k] = \ol Z_k Z_k - Z_k \ol Z_k =
                          \Cov_{\ol Z_k} Z_k - \Cov_{Z_k} \ol Z_k$.
\end{proof}

\begin{rem}
  If the curvature $R$ of the bundle and the K\"ahler 
form $\omega =\frac{1}{2} g(\cdot,J\cdot)$
are in the prequantum relation 
\begin{equation}
\label{eq:pqrel}
  R_{X,Y}=\frac i \hbar \omega(X,Y) 
\end{equation}
for any $X,Y \in \Upsilon(\M)$,
then $\rho$ is a constant, 
\begin{gather}
    \rho = i \sum_{k=1}^{d/2} R_{E_k,JE_k} 
    = - \frac{1}{2\hbar} \sum_{k=1}^{d/2} g(E_k,E_k) = - \frac{d}{4\hbar} \, .
\end{gather}
\end{rem}

\subsection{Berezin-Toeplitz Operators as Limits of  Schr\"o\-din\-ger
         Operators}

This subsection shows how a Berezin-Toeplitz operator can be extended
to a family of Schr\"o\-din\-ger operators and be reconstructed as a monotone 
limit of this family. A major benefit is that the knowledge about 
Schr\"o\-din\-ger operators may be used to find sufficient conditions 
for the semiboundedness of\/ $\t_f$, thereby ensuring the self-adjointness
of the associated Berezin-Toeplitz operator.

\begin{convention}
In the following, $\L$ is always a holomorphic Hermitian line bundle and
$g$ is assumed to be the real part of a K\"ahler metric 
on the $d$-dimensional base manifold $\M$. 
\end{convention}

\begin{prop} \label{prop:SqSoiq}
If the manifold $\M$ is complete, then
the space $L^2_\hol(hm)$ is in the domain of the 
form-closure of $\cE^{(0,\bullet)}$ and can be identified
as the null-space $\{\psi \in L^2(hm): -\Delta^{(0,\bullet)} \psi = 0\}$
of the holomorphic Laplacian $-\Delta^{(0,\bullet)}$.    
\end{prop}
\begin{proof}
  Given $\psi \in L^2_\hol(hm)$, we need to construct a Cauchy sequence 
$\{\psi_l\}_{l \in \N}$
in $\Cinfty_{c\L}(\M)$ which converges to $\psi$ with respect to
the form-norm, $\norm{\psi_l-\psi}_{\cE^{(0,\bullet)}} \to 0$.
To this end, we use an increasing sequence of localizing
cut-off functions $\eta_l: \M \to [0,1]$ observing the uniform
gradient bound $\sup_{x \in \M} \norm{\grad \eta_l (x)} \leq \frac C {2^l}$
for some constant $C>0$, as described in the preceding part of this section.
% cite Whitney
Then by monotone convergence
$\norm{\eta_l \psi - \psi} \to 0$, and the remaining term in  the form-norm
can be estimated by
\begin{eqnarray}
  \label{eq:Bginformclosure}
  \cE^{(0,\bullet)}(\eta_l \psi,\eta_l \psi) 
   &=& \int_\M \sum_k h(\nabla_{\ol Z_k} \eta_l \psi, \nabla_{\ol Z_k} \eta_l \psi) \, dm \\
 &=& \sum_k \int_\M \Bigl(\abs{Z_k(\ol \eta_l)}^2 h(\psi, \psi) + 
                 \abs{\eta_l}^2  
                 h(\nabla_{\ol Z_k} \psi, 
                   \nabla_{\ol Z_k} \psi ) \,  \Bigr. \nonumber \\
  && \Bigl. \phantom{\sum\int\Bigl(\Bigr.} + 2 \, \Re \bigl(Z_k(\ol \eta_l) 
            h ( \psi, \nabla_{\ol Z_k} \psi ) \bigr) \Bigl) \, dm \\
  & \leq & \frac{C^2}{2^{2l}} \norm{\psi}^2
          + \int_\M \abs{\eta_l}^2  \sum_k 
                h(\nabla_{\ol Z_k} \psi, \nabla_{\ol Z_k} \psi) \, dm \nonumber \\
 &&+ 2 \int_\M \sum_k \abs{Z_k(\ol\eta_l) h(\psi, \nabla_{\ol Z_k} \psi ) } \, dm \, .
\end{eqnarray}
Using the Cauchy-Schwarz inequality, we have
\begin{equation}
  \label{eq:BginformclII}
   \cE^{(0,\bullet)}(\eta_l \psi,\eta_l \psi) 
\leq \frac{C^2}{2^{2l}} \norm{\psi}^2 + \cE^{(0,\bullet)}(\psi,\psi)
    + \frac{2C}{2^l} \norm{\psi} (\cE^{(0,\bullet)}(\psi,\psi))^{1/2} 
\end{equation}
so by dominated convergence
$\cE^{(0,\bullet)}(\eta_l \psi - \psi, \eta_l \psi - \psi) \to 0$. 
Thus, both terms in the form-norm converge to zero.
\end{proof}

\begin{defn} \label{def:Schrop} 
  A semibounded Schr\"o\-din\-ger operator $S^\cL_{D,q}$ on $L^2(hm)$
is the self-adjoint operator associated with the form
\begin{equation}
  \label{eq:Sop}
  \cS^\cL_{D,q}(\psi,\psi) = D \cE^\cL(\psi,\psi) + (\psi, q \psi) \, ,
\end{equation}
where $D>0$ is some coupling constant and
the requirement 
\begin{equation}
  \label{eq:formbd}
  (\psi,q^-\psi) \leq c_1 \cS^\cL_{D,q^+}(\psi,\psi) + c_2 (\psi,\psi)
\end{equation}
is satisfied
with relative form bound $c_1 < 1$ and some constant $c_2 \ge 0$.
Thus, the form domain of $\cS^\cL_{D,q}$ is
obtained from the closure of $\Cinfty_{c\L}(\M) \cap 
\{ \psi: (\psi,q^+\psi) < \infty\}$.
\end{defn}

\begin{rem} \label{rem:S01Df}
If in addition to the requirement \eq{formbd}
the curvature term $\rho$ of Proposition~\ref{prop:WBF} 
is also a form-bounded perturbation of $S^\cL_{q^+}$, then 
  \begin{equation}
    \label{eq:S01Df}
\cS^{(0,\bullet)}_{D,q}: \psi \mapsto \cS^\cL_{D,D\rho + q}(\psi,\psi) 
                                     = D \cE^{(0,\bullet)}(\psi,\psi) + (\psi,q\psi)  
  \end{equation}
has the same form domain as $\cS^\cL_{D,q}$ and
also defines a generalized Schr\"o\-din\-ger operator, 
hereafter referred to as $S^{(0,\bullet)}_{D,q}$.
\end{rem}

\begin{thm}
If the assumptions of Proposition~\ref{prop:SqSoiq} 
and Remark~\ref{rem:S01Df}
are fulfilled, 
then $\cS^{(0,\bullet)}_{D,f}$ is 
semibounded and $\t_f$ on $L^2_\hol(h\meas)$
is closed and semibounded. 
\end{thm}
\begin{proof}
First we note $\cS^{(0,\bullet)}_{D,f}(\psi,\psi) = \t_f(\psi,\psi)$ for any $D>0$ and 
$\psi \in \cQ(\t_f)$. 
Thus, we only need to show that
the restriction of $\cS^{(0,\bullet)}_{D,f}$ to the closed subspace 
$L^2_\hol(hm)$ is again a closed and semibounded form. 

  To show closedness, assume a sequence $(\psi_l)_{l \in \mathbb N}$  
in $L^2_\hol(hm)$
which is Cauchy with respect to the form-norm. Then by the closedness of 
$\cS^{(0,\bullet)}_{D,f}$
the sequence has a limit $\psi \in \cQ(\cS^{(0,\bullet)}_{D,f})$. However, this limit 
is contained
in $L^2_\hol(hm)$, because
the sequence $(\psi_l)_{l \in \mathbb N}$ also  converges
with respect to the usual norm on $L^2_\hol(hm)$.

Semiboundedness follows from the inequality
\begin{equation}
  \label{eq:lbd}
  \inf_{\substack{\psi \in L^2(hm),\\ \norm{\psi} = 1}} \cS^{(0,\bullet)}_{D,f}(\psi,\psi)
  \leq \inf_{\substack{\psi \in L^2_{\mathit{hol}}(hm),\\ \norm{\psi} = 1}} 
                      \cS^{(0,\bullet)}_{D,f}(\psi,\psi)  \, 
\end{equation}
due to the set inclusion $L^2_{\mathit{hol}}(hm) \subset L^2(hm)$.
\end{proof}

\begin{remarks} \label{rem:notdense} 
As stated, the above theorem does not imply that $\t_f$ 
is densely defined. 
Therefore, $T_f$ might be self-adjoint only on a Hilbert-subspace of $L^2_\hol(hm)$.

In analogy with Lemma~\ref{thm:KLMN}, it is sufficient for the closedness 
and semiboundedness of $\t_f$ when for some $D>0$ the negative part
$f^-$ can be incorporated as a form-bounded perturbation of $\cS^{(0,\bullet)}_{D,f^+}$
with relative form bound strictly less than one.
However, this condition is not as easy to characterize in terms of $f$ 
as the stronger assumption in the preceding theorem.
\end{remarks}

\begin{consequence} \label{prop:sgcv}
If the assumptions of the preceding theorem hold, then
the semigroup generated by $S^{(0,\bullet)}_{D,f}$
converges in the limit $D \to \infty$ strongly
to a Berezin-Toeplitz semigroup, 
  \begin{equation}
    \label{eq:sgconv}
    \lim_{D \to \infty} \ep{-t  S^{(0,\bullet)}_{D,f}} \psi = \ep{- t T_f} 
                                            K_f \psi ,
  \end{equation}
where $t>0$, $\psi \in L^2(hm)$, and the orthogonal projector $K_f= K_f^* K_f$ maps 
onto the closure of
$\cQ(\t_f)$ in $L^2_\hol(hm)$.
\end{consequence}
\begin{proof}
The limit $D \to \infty$ of 
$\cS^{(0,\bullet)}_{D,f}$ yields a non-densely defined quadratic form 
\begin{equation}
  \label{eq:sinfty}
    \cS^{(0,\bullet)}_{\infty,f} : 
     \psi \mapsto \lim_{D \to \infty} \cS^{(0,\bullet)}_{D,f}(\psi,\psi)  
\end{equation}
which is by inspection identical with $\t_f$.

The monotone convergence implies then that $\t_f$ is closed \cite{Sim78} 
and the semiboundedness follows
from that of $\cS^{(0,\bullet)}_{D,f}\le \t_f$ for some $D>0$.
These properties imply that the Berezin-Toeplitz operator $T_f$ associated 
with $\t_f$ is self-adjoint on the closure $\ol{\cQ(\t_f)} \subset L_\hol^2(hm)$.
Again by the monotone convergence of forms 
the self-adjoint operators $S^{(0,\bullet)}_{D,f}$
converge in the strong resolvent sense \cite{Sim78}, 
which in turn implies strong convergence of the semigroups they generate 
\cite[Theorem S.14]{RS80}. 
\end{proof}

%%%%%%%%%%%%%%%%%%%%%%%%%%%%%%%%%%%%%%%%
% Example of BGV Prop 2.5
%%%%%%%%%%%%%%%%%%%%%%%%%%%%%%%%%%%%%%%%

% ----------------------------------------------------------------------
\section{Probabilistic Representation of Berezin-Toeplitz Semigroups}
\label{ch:5}
% ----------------------------------------------------------------------

This section introduces a new element into the discussion
of Berezin-Toeplitz operators, 
the concept of Brownian motion on the base manifold $\M$ of the holomorphic
line bundle $\L$.
In terms of this stochastic process, one may characterize the Kato class
of functions on $\M$. It turns out that Kato decomposable functions
$f$ lead to semibounded, self-adjoint Berezin-Toeplitz operators
$T_f$ on $L^2_\hol(hm)$, where $m$ is the Riemannian \mbox{volume} measure
on $\M$. The final result in this section is a probabilistic 
expression for Berezin-Toeplitz semigroups, referred to as the 
Daubechies-Klauder formula.
It is convenient that the validity of the Daubechies-Klauder formula
is expressed in terms of the Kato class associated with Brownian
motion and thus the admissibility of bundle curvature and classical Hamiltonian
in the quantization procedure and path-integral formulation are 
intrinsically characterized by the underlying geometry.

\subsection{The Kato Class and a Version of the Feynman-Kac Formula}

We adopt the usual terminology: 
An amost surely continuous process $\sB$ with values in 
the Riemannian manifold $\M$ is called Brownian
motion with diffusion constant $D>0$
if for every smooth function $\phi \in \Cinfty(\M)$, the difference
\begin{equation}
  \label{eq:BM}
  \sM_t := \phi\circ \sB_t - \phi \circ \sB_0 -  \int_0^t D \Delta \phi \circ \sB_s ds
\end{equation}
is a real-valued continuous local martingale $\sM$.

A probability measure governing Brownian motion with 
diffusion constant $D>0$ and almost
surely fixed starting point $\sB_0=x \in \M$
will be denoted as $\bP^D_x$. The expectation
with respect to this probability measure
is written as $\E^D_x$.

  A Riemannian manifold $\M$ is called Brownian complete if for
a fixed diffusion constant $D>0$, 
a Brownian motion $\sB$ starting at any $x \in \M$
has an infinite explosion time.   

%We denote by $\{p_t\}$ and $\{p^\cL_t\}$ the unique 
%heat kernels associated with the semigroups generated
%by the Laplacians $\Delta$ and $\Delta^\cL$.

\begin{defn}
Let $\M$ be a Brownian-complete Riemannian manifold with
a family of Brownian-motion measures $\fam{\bP^D_x}{x \in \M}$ 
having a fixed diffusion constant $D>0$. 
A real-valued function $q: \M \to \R$ belongs to the
Kato class $\Kato$ if the following 
condition is satisfied:
\begin{equation}
  \label{eq:K_def}
  \lim_{t \downto 0} \sup_{x \in \M} \int_0^t \E^D_x[\abs{q}(\sB_s)] \, ds = 0 \; .
\end{equation}
Whenever this property holds only locally, which means for all products
$\chi_\Lambda q \in \Kato$ with characteristic functions $\chi_\Lambda$
of compact sets $\Lambda$ in $\M$, we write $q \in \Kloc$. 
If a real-valued function $q$ satisfies $q^+ \in \Kloc$ and $q^- \in \Kato$ 
then it is called Kato decomposable, symbolized as $q \in \Katopm$.  
\end{defn}

\begin{itremarks}
If a function has the global or local Kato property for one choice of $D>0$, then
this holds for any $D>0$. The reason to include $D$ in the definition is merely
for the consistency of notation. 

If the Ricci curvature of a Riemannian manifold is bounded from below,
then the kernel is on compact sets up to any finite time
uniformly bounded away from zero \cite{Dav89,Stu92}.
As a consequence, the local Kato property implies local integrability
with respect to the volume measure, $\Kloc \subset L^1_\loc(m)$. 
%form defined on C_c\infty
\end{itremarks}

The following lemma goes back to Kha\'sminskii \cite{Kha59}. Our discussion
of the Kato class proceeds along the nice exposition in 
\cite[Section 1.2]{Szn98}.
The sole purpose of the following passage is to show that
functions from the Kato class can be viewed as infinitesimally form-bounded
perturbations of the Dirichlet and Bochner Laplacians.

\begin{lem} %Kashminskii
\label{lem:khas} 
  Suppose $0 \leq q \in \Kato$, and $t > 0$ is chosen such that
  \begin{gather}
    \label{eq:KhasI}
    \kappa := \sup_{x \in \M} \E^D_x\left[ \int_0^t q(\sB_r) dr \right] < 1 \, , 
\intertext{then} 
  \label{eq:KhasII}
  \sup_{x \in \M} \E^D_x\left[ \ep{ \int_0^t q(\sB_r) dr } \right] \leq \frac{1}{1 - \kappa} \, .
\end{gather}
\end{lem}
\begin{proof} We refer to the proof in \cite[Lemma 2.1]{Szn98}, which is
concerned with the special case of Brownian motion in Euclidean
space. However, the formulation given there transfers literally without
modification to the general situation on manifolds. The essential steps are
to expand the exponential, to use time-ordering and the Markov property,
and to inductively apply the assumption to obtain a geometric series in 
$\kappa$.
\end{proof}

\begin{consequence}
Kha\'sminskii's lemma implies that for $q \in \Kato$,
the mapping $Q_t$ given by
\begin{equation} \label{eq:Qtphi}
  Q_t\phi(x) := \E^D_x\Bigl[ \ep{ \int_0^t q(\sB_r) dr } \phi(\sB_t)\Bigr] 
\end{equation}
has a bound $\norm{Q_t}_{\infty,\infty} 
:= \sup_{x \in \M, \norm{\phi}_\infty=1} \abs{Q_t \phi(x)}\leq e^{Ct}/(1-\kappa)$ 
on $L^\infty(m)$ with exponential growth in $t$.
Thus $\{Q_t\}_{t\ge 0}$ is a semigroup of operators on $L^\infty(m)$.
\end{consequence}
\begin{proof}
Again, we refer to \cite[Theorem 2.2]{Szn98}. The idea is to split
$[0,t]$ into subintervals, to inductively use Lemma~\ref{lem:khas} in 
conjunction with the Markov property, and to bound the resulting 
expression with an exponential. 
\end{proof}

\begin{lem}
  Given a non-negative function $q \in \Kato$, then for any 
$c_1>0$ there is a $c_2>0$ such that
\begin{equation}
  \label{eq:fbdcond}
  \int_\M q \abs{\phi}^2 dm \leq c_1 D \cE(\phi,\phi) + c_2 \norm{\phi}_2^2
\end{equation}
whenever $\phi \in \cQ(\cE)$. In other words,
functions from the Kato class act as infinitesimally form-bounded
perturbations of $-D\Delta$.
\end{lem}
\begin{proof}
  The proof proceeds in two steps.\\
\textit{Step 1.} For $q \in \Kato$, the expression \eq{Qtphi}
defines a strongly continuous semigroup $\fam{Q_t}{t\ge 0}$ of bounded, self-adjoint
operators $Q_t$ on $L^2(m)$.

The preceding lemma together with the Jensen and Cauchy-Schwarz inequalities 
establish
the boundedness,
\begin{align}
 & \int_\M \abs{ \E_x^D[\ep{\int_0^t q(\sB_s)ds} \phi(\sB_t)]}^2 dm(x) \\
 & \leq \int_\M \E_x^D[\ep{2 \int_0^t q(\sB_s)ds}] \E_x^D[ \abs{\phi(\sB_t)}^2] dm(x)
  \leq \frac{1}{1-\kappa} e^{Ct} \norm{\phi}^2_2 \, .
\end{align}
Because of the Markovian semigroup property,
it is enough to show strong continuity at $t=0$. To this end, we consider
\begin{align} \label{eq:l2norm}
 & \lim_{t \downto 0} \int_\M \abs{ \E_x^D[(\ep{\int_0^t q(\sB_s)ds} -1)\phi(\sB_t)]}^2 dm(x) \\
 & \leq  \lim_{t \downto 0}  \int_\M \E_x^D[(\ep{\int_0^t q(\sB_s)ds}-1)^2] 
                   \E_x^D[ \abs{\phi(\sB_t)}^2] dm(x) \\
 & \leq \lim_{t \downto 0} \sup_{x \in \M} \E_x^D[e^{2 \int_0^t \abs{q}(\sB_s)ds} -1]
                           \int \E^D_x[\abs{\phi(\sB_t)}^2] dm(x) \, .
\end{align}
The last step involves H\"older's inequality and the elementary estimate
$(e^c-1)^2 \leq e^{2\abs c} - 1$ for any real number $c \in \R$.
Using the definition of the Kato class in this estimate shows that
 the limit of 
the $L^2$-norm in \eq{l2norm} vanishes.

Moreover, by the time reversal invariance of Brownian motion
each $Q_t$ is seen to be self-adjoint, and
according to the Hille-Yosida theorem, there is a semibounded, self-adjoint
generator of the semigroup.
Given any $c_1 > 0$, we choose a suitable constant $c_2$ 
such that replacing $q$ by
$\tilde q := q/c_1 - c_2/c_1$ in the above procedure 
yields a contraction semigroup.\\
\textit{Step 2.} 
If we approximate $\tilde q \in \Kato$ with a sequence
of semibounded functions $\tilde q_l := \min\{ \tilde q , l\}$, 
then for any $\phi \in \cQ(\cE)$
the generator of the contraction semigroup $\widetilde Q_t$ 
associated with the function $\tilde q_l$
gives rise to a quadratic form
\begin{align}
  \lim_{t \downto 0} \frac 1 t (\phi, \widetilde Q_t \phi - \phi)
  = - D \cE(\phi,\phi) -  \frac{c_2}{c_1} \norm{\phi}_2^2 
                       +  (\phi,\min\{\frac{q}{c_1},l+\frac{c_2}{c_1}\} \phi) \leq 0
\end{align}
because $-D \Delta$ is essentially self-adjoint and the multiplication by 
$\tilde q_l$ is a bounded operator. 
The contractivity of the semigroup 
furnishes the last inequality, which in turn
yields the form-boundedness condition 
\eq{fbdcond} by monotone convergence in the limit $l \to \infty$ .
\end{proof}

\begin{prop}
Let $\L$ be a Hermitian line bundle with a connection
and an associated length-preserving horizontal transport $H$. Suppose
the base  manifold
$\M$ is Riemannian and Brownian complete, equipped with a
family of Brownian motion measures $\{\bP_x^D\}_{x \in \M}$ having a 
common diffusion constant $D>0$.
Then $q \in \Kato$ is also a form-bounded perturbation
of the negative Bochner Laplacian $-\Delta^\cL$. 
\end{prop}
\begin{proof}
To make contact with the preceding lemma,
we fix $\psi \in L^2(hm)$ and define a function
$\phi \in L^2(m)$ with values $\phi(x) := \sqrt{h_x(\psi(x),\psi(x))}$.

We may now verify the $L^2$-boundedness
of the operators $Q^\cL_t$ given by
\begin{equation}
  Q^\cL_t \psi(x) := \E_x^D\Bigl[ \ep{\int_0^t q(\sB_r) dr}H^{-1}_{\sB,t}\psi(\sB_t)\Bigr]
\end{equation}
with an estimate using that horizontal transport preserves the Hermitian metric
and the same strategy as in the preceding lemma,
\begin{align}
  \sqrt{h_x(Q^\cL_t\psi(x),Q^\cL_t\psi(x))} 
 &\leq \E^D_x\left[ \ep{\int_0^t q(\sB_r) dr} \sqrt{ h_x(H^{-1}_{\sB,t}\psi(\sB_t),
                                                  H^{-1}_{\sB,t}\psi(\sB_t))}\right]\\
 &=  \E^D_x\left[ \ep{\int_0^t q(\sB_r) dr} \sqrt{h_{\sB_t}(\psi(\sB_t),\psi(\sB_t))}\right]\\
 &= \E^D_x\left[ \ep{\int_0^t q(\sB_r) dr} \phi(\sB_t) \right]
      = e^{-tS_{D,q}}\phi(x) \, .
\end{align}
A similar estimate gives
\begin{align}
  (\psi, Q_t^\cL \psi - \psi) \leq (\phi, Q_t \phi - \phi) 
\end{align}
and thus together with the preceding lemma the desired form-boundedness.
\end{proof}

\begin{itconsequence}
  Therefore, any $q \in \Kato$ may be used to define a form-bounded
perturbation of $-D\Delta$ or $-D\Delta^\cL$ in order to define
a self-adjoint Schr\"odinger operator.
One may also use $\cS_{D,q^+}$ or $\cS^\cL_{D,q^+}$ as the unperturbed
forms and thus extend this construction to define a Schr\"odinger operator
$\cS_{D,q}$ with $q \in \Katopm$.
\end{itconsequence}

\begin{fk}
Let $\L$ be a Hermitian line bundle with a connection
and a length-preserving horizontal transport $H$. Suppose
the base  manifold $\M$ is Riemannian and Brownian complete. 
 Denote by $m$ the natural volume measure on $\M$
and by $\bP^D$ a family of Brownian-motion measures 
having a common diffusion constant $D>0$.

If $q \in \Katopm$, then the
image of a section $\psi \in L^2(hm)$ under the 
semigroup $e^{-tS^\cL_{D,q}}$ generated by
the self-adjoint Schr\"odinger operator $S^\cL_{D,q}$ has the
probabilistic representation
  \begin{equation}  \label{eq:FKII}
    e^{-tS^\cL_{D,q}} \psi(x)
 = \E^D_x\Bigl[ e^{-\int_0^t q(\sB_r) dr} H^{-1}_{\sB,t} \psi(\sB_t) \Bigr] \, .
  \end{equation}
The inverse of the stochastic horizontal transport appearing in this equation
can either be understood by appealing to localized expressions \cite{Sch80},
i.e.\ one restricts to a subspace of the probability space 
by introducing exit times of local coordinate patches and then reformulates
the horizontal transport in a local trivialization, or one interprets 
\eq{FKII} as a shorthand for
\begin{equation}
  \label{eq:heatLeqnII}
  h_x(u,e^{-tS^\cL_{D,q}} \psi(x))
 = \E^D_x\left[e^{-\int_0^t q(\sB_r) dr} h_{\sB_t}(H_{\sB,t} u,\psi(\sB_t))\right] \,  
\end{equation}
with an arbitrary reference vector $u \in \L_x$.
\end{fk}
\begin{proof}
  The proof of formula \eq{FKII} is given in Appendix~\ref{app:ItoLB}.
\end{proof}

\begin{consequence} \label{con:I}
Let $\L$ be a holomorphic Hermitian line bundle, assume its base manifold
$\M$ is K\"ahler, and let $g$ denote the real part of the K\"ahler metric.
Let $\{Z_k\}_{k =1}^{d/2}$ denote a local holomorphic orthonormal 
frame of $T\up{1,0}\M$.
If the curvature term $\rho = \sum_{k=1}^{d/2} R_{\ol Z_j,Z_j}$
determined by the connection of the line bundle $\L$
is Kato decomposable, then $-\Delta^\cL$ and 
$-\Delta^{(0,\bullet)}$ have the same domain and are essentially self-adjoint 
on $\Cinfty_{c\L}(\M)$. If $f$ is also Kato decomposable, then
the Feynman-Kac formula 
\eq{FKII} with $q=D\rho+f$  
gives an expression for the Schwartz kernel of the 
semigroup generated by $S_{2D,f}^{(0,\bullet)}$.
\end{consequence}

\begin{defn}
With the help of the distance function on $\M$, the space $C_\M([0,t])$
of continuous paths in $\M$ parametrized by an interval $[0,t]$
can be turned into a complete, separable metric space.
In this setting, one may construct a regular conditional probability
measure of $\bP^D_x$ given $\sB_t$ \cite[Theorem 5.3.19]{KS91}.
We denote by $\bP_{x,y}^{D,t}$ the Brownian bridge measure,
that is the probability measure of the Brownian motion which starts at $\sB_0=x$
and is conditioned to arrive at $\sB_t=y$. It is understood as a regular 
conditional probability distribution of $\bP_x^D$ given $\sB_t$ 
\cite[pp.~146--150]{Par67}.
\end{defn}

\begin{consequence} \label{con:PR}
With the assumptions of Consequence~\ref{con:I},
the Feynman-Kac formula \eq{FKII} can be modified to give an expression for the
integral kernel of the Schr\"o\-din\-ger semigroup 
\begin{equation}
  \label{eq:FKkernel}
    \ep{-t S^{(0,\bullet)}_{2D,f}}(x,y)
       =  p_{D,t}(x,y) \E^{D,t}_{x,y}\Bigl[ \ep{-\int_0^t (D\rho(\sB_r) + f(\sB_r)) dr} 
                                            H^{-1}_{\sB,t} \Bigr] 
\end{equation}
generated by $S^{(0,\bullet)}_{2D,f}$.
Hereby, we use the unique heat kernel $\{p_{D,t}\}_{t > 0}$
of $D\Delta$, the 
expectation with respect to the Brownian bridge 
measure $\bP_{x,y}^{D,t}$ given above, and 
the inverse of the stochastic horizontal transport is
understood as a linear mapping from $\L_y$ to $\L_x$. 
\end{consequence}

\begin{defn}
In the following results, we always consider a fixed   
Kato-decompo\-sab\-le symbol $f \in \Katopm$. To simplify notation, we choose
a reference diffusion constant $D_0 > 0$ 
and abbreviate for $c \ge 0, v \in \L$
the section
\begin{equation}
    \label{eq:etatxy}
       \eta_{c,v} := \ep{- S^{(0,\bullet)}_{2D_0,c f}}(\cdot,\pi(v))v 
\end{equation}
obtained by keeping one end of the Schwartz kernel fixed.
\end{defn}

\begin{lem}
 For any $v \in \L$ and $c \ge 0$, the section $\eta_{c,v}$
is contained in $L^2(hm)$. In addition, the mapping 
$c \mapsto \eta_{c, v}$ is strongly continuous.
\end{lem}
\begin{proof}
  Due to the linearity in $v$, it is enough to consider a vector of length $\norm{v}=1$.
The $L^2$-norm of $\eta_{c, v}$ can then be estimated by repeatedly using 
the Cauchy-Schwarz and H\"older inequalities:
\begin{align}
  \norm{\eta_{c, v}}_2 
        &\leq \sup_{\substack{v\in \L_y, \norm{v}=1 \\ 
                             \psi \in L^2(hm), \norm{\psi}_2=1}} \!\!
                    \abs{(\eta_{c, v},\psi)} 
        \leq \sup_{\substack{v \in \L, \norm{v}=1 \\
                     \psi \in L^2(hm), \norm{\psi}_2=1}} \!\!
                         \Bigl| h_{\pi(v)} (e^{- S^{(0,\bullet)}_{2D_0,c f}} \psi (\pi(v)),v)\Bigr| \\
 %      &=& \Bnorm{e^{-t S^{(0,\bullet)}_{D,c f}}}_{2,\infty}\\
        &\leq \sup_{\substack{\norm\psi_2 =1, x \in \M}} 
                      \hnorm{ \E_{x}^{D_0} \left[ \ep{-\int_0^1 (D_0\rho +c f)(\sB_t) dt} H^{-1}_{\sB,1} \psi(\sB_1) \right] }\\
       &\leq \sup_{\substack{\norm\psi_2 =1, x \in \M}} 
                 \left( \E_{x}^{D_0} \left[ \ep{\mbox{\,\footnotesize $-2$}\!\int_0^1 (D_0\rho + c f)(\sB_t) dt} \right] 
                        \E_{x}^{D_0} \left[ \norm{\psi(\sB_t)}^2 \right] \right)^{1/2}\\
       &\leq \Bigl(\Bnorm{ \ep{-S^{\L}_{D_0,2D_0\rho + 2c f}} }_{\infty,\infty}\Bigr)^{1/2}
                  \norm{p_1(\centerdot,x)}_\infty < \infty \, .
\end{align}
The finiteness results from the Kato decomposability of $\rho$ and $f$ and
from the boundedness of the heat kernel $p_1$ \cite{Dav88,Dav89,Stu92}.

To see the strong continuity of $\eta_{c,v}$ in $c$,
we may for cimplicity assume $v$ to be again a normalized vector
and consider two non-negative coupling constants $c$ and $c'$.
If $\psi \in L^2(hm)$ also has the $L^2$-norm $\norm{\psi}_2=1$, then
we may estimate
\begin{align}
   &\abs{(\eta_{c,v} - \eta_{c',v},\psi)}  =
    \Bigl| \E^{D_0}_{\pi(v)}\Bigl[ \Bigl(
                          \ep{-\int_0^1(D_0\rho + c f)(\sB_t)dt} \Bigr.\Bigr.\Bigr.\nonumber \\
&\phantom{  =\E^{D_0}_{\pi(v)}\Bigl[\Bigr]==========}  
   \Bigl.\Bigl.\Bigl. - \ep{-\int_0^1(D_0\rho + c' f)(\sB_t)dt} \Bigr)
                          h_{\sB_1}(H_{\sB,1} v, \psi(\sB_1)) \Bigr] \Bigr| \nonumber \\
& \leq   \Bigl(\E^{D_0}_{\pi(v)}\bigl[ \abs{ h_{\sB_1}(v,\psi(\sB_1))}^2 \bigr]\Bigr)^{1/2}
         \Bigl(   \E^{D_0}_{\pi(v)}\Bigl[ \Bigl(
                          \ep{-\int_0^1(D_0\rho + c f)(\sB_t)dt} \Bigr.\Bigr.\Bigr.\nonumber \\
  &\phantom{  \E^{D_0}_{\pi(v)}\bigl[ h_{\sB_1}(\psi(\sB_1),\psi(\sB_1)) \bigr]  
              \E^{D_0}_{\pi(v)}\Bigl[ \Bigl( \Bigr)\Bigr] ===}  
   \Bigl.\Bigl. - \ep{-\int_0^1(D_0\rho + c' f)(\sB_t)dt} \Bigr)^2\Bigr]\Bigr)^{1/2} \\
& \leq   \Bigl(\E^{D_0}_{\pi(v)}\bigl[ h_{\sB_1}(\psi(\sB_1),\psi(\sB_1)) \bigr]\Bigr)^{1/2}  
         \Bigl(\E^{D_0}_{\pi(v)}\Bigl[ \Bigl(
                          \ep{-\int_0^1(D_0\rho + c f)(\sB_t)dt} \Bigr.\Bigr.\Bigr.\nonumber \\
  &\phantom{  \E^{D_0}_{\pi(v)}\bigl[ h_{\sB_1}(\psi(\sB_1),\psi(\sB_1)) \bigr]  
              \E^{D_0}_{\pi(v)}\Bigl[ \Bigl( \Bigr)\Bigr] ====}  
   \Bigl.\Bigl.\Bigl. - \ep{-\int_0^1(D_0\rho + c' f)(\sB_t)dt} \Bigr)^2\Bigr] \Bigr)^{1/2}\, .
\end{align}
Taking the supremum over $\psi, \norm{\psi}_2=1$ on both sides 
together with the $L^2$-con\-trac\-tion property of the
unperturbed heat semigroup generated by $D_0\Delta^\cL$ yields
\begin{align}
  &\norm{ \eta_{c, v} - \eta_{c', v}}_2
 = \sup_{\psi \in L^2(hm), \norm{\psi}_2 = 1} 
           \abs{(\eta_{c,v} - \eta_{c',v},\psi)} \nonumber \\
% &= \sup_{\psi \in L^2(hm), \norm{\psi}_2 = 1} 
%      \abs{\E^{D_0}_{\pi(v)}\Bigl[ \Bigl(
%                          \ep{-\int_0^1(D_0 + c f)(\sB_t)dt}
%                        - \ep{-\int_0^1(D_0 + c' f)(\sB_t)dt} \Bigr)
%                          H^{-1}_{\sB,1} \psi(\sB_1) \Bigl]} \\
% &\leq  \sup_{\psi \in L^2(hm), \norm{\psi}_2 = 1} 
%        \E^{D_0}_{\pi(v)}\Bigl[ \Bigl(
%                          \ep{-\int_0^1(D_0 + c f)(\sB_t)dt}
%                        - \ep{-\int_0^1(D_0 + c' f)(\sB_t)dt} \Bigr)^2\Bigl]
%         \E^{D_0}_{\pi(v)}\bigl[ h_{\sB_1}(\psi(\sB_1),\psi(\sB_1)) \bigr] \\           
 &\leq  \Bigl(\E^{D_0}_{\pi(v)}\Bigl[ \Bigl(
                          \ep{-\int_0^1(D_0\rho + c f)(\sB_t)dt}
                        - \ep{-\int_0^1(D_0\rho + c' f)(\sB_t)dt} \Bigr)^2\Bigl] \Bigr)^{1/2}
          \label{eq:DC}\\
 &\leq 2  \Bigl(\E^{D_0}_{\pi(v)} \Bigl[ \ep{2\int_0^1(D_0\rho^- + c_0 f^-)(\sB_t)dt}
                            \Bigr] \Bigr)^{1/2}\, .
\end{align}
The purpose of the last estimate is to show that with the help of some
large $c_0$, dominated convergence applies to \eq{DC}
in the limit $c'\to c$. 
\end{proof}

\begin{dk}
 Let $\L$ be a holomorphic line bundle with a Hermitian metric $h$,
suppose its base manifold $\M$ is equipped with a K\"ahler metric, 
denote its real part by $g$ and the natural volume measure by $m$. 
Let $\M$ be Riemannian complete with Ricci curvature bounded
below, to ensure Brownian completeness.
Let the real-valued function $f: \M \to \R$ be Kato decomposable 
with respect to the Brownian-motion measure $\bP^D$ on $\M$,
where the diffusion constant $D>0$ is arbitrary. 
In addition, suppose the curvature term $\rho$
defined in Proposition \ref{prop:WBF} is also Kato decomposable.
To include the case when $T_f$ is not densely defined, we 
denote by $K_f$ the orthogonal projector onto the 
closure of the form domain $\cQ(\t_f)$ in $L^2_\hol(hm)$.

With these assumptions the integral kernel of the
Berezin-Toeplitz semigroup $\{e^{-tT_f}K_f\}_{t\ge 0}$ is for $t>0$ 
given by the probabilistic expression
\begin{equation}
  \label{eq:DKformula}
  \bigl(e^{-tT_f}K_f\bigr)(x,y) = \lim_{D \to \infty} 
     p_{D,t}(x,y) \E_{x,y}^{D,t}\Bigl[ \ep{-\int_0^t (D\rho(\sB_r) + f(\sB_r))dr}
                                                      H^{-1}_{\sB,t} \Bigr] \, .
\end{equation}
 In particular, we obtain the reproducing kernel $K$ of $L^2_\hol(hm)$
as a special case of this formula when $f=0$.
\end{dk}
\begin{proof}
In conjunction with the specific use of the Feynman-Kac formula
in Consequence~\ref{con:PR}, it is equivalent to show that
the integral kernel of the Berezin-Toeplitz semigroup $\ep{-t T_f}K_f$ 
on $L^2_\hol(hm)$ is for $t>0$ given by the pointwise limit 
\begin{equation}
  \label{eq:DKsgform}
  \bigl(\ep{-tT_f}K_f\bigr)(x,y) = \lim_{D \to \infty} \ep{-t S^{(0,\bullet)}_{D,f}}(x,y) \, ,
\end{equation}
where $S\up{0,\bullet}_{D,f}$ is the semibounded Schr\"odinger operator
defined by equation \eq{S01Df}.

We borrow the strategy of \cite{BLW99} and  
accommodate it to the manifold situation and the case of unbounded $f$.
The key to the present generalization is the use of monotone form convergence.

We have to show that for $u,v \in \L$ with base points $x, y \in \M$
the equation
\begin{equation}
  \label{eq:etatok}
    \lim_{D \to \infty} h_x(u,\ep{-t S\up{0,\bullet}_{D,f}}(x,y)v) 
       = (e_u, \ep{-t T_f} K_f e_v)
\end{equation}
holds, which by Consequence~\ref{con:intkernels}  
characterizes the continuous integral kernel of 
$e^{-t T_f} K_f$ on $L^2_\hol(hm) \subset L^2(hm)$.

To see \eq{etatok},
we use the semigroup property and express the integral kernel 
with some choice of $D_0 >0$ in a scalar product 
\begin{equation}
  \label{eq:eee}
  h_x(u,\ep{-t S\up{0,\bullet}_{D,f}}(x,y)v) 
      = \Bigl(\eta_{D_0/D, u}, 
          \exp\bigl(-t S^{(0,\bullet)}_{D - 2 D_0, (D - 2 D_0)f/D}\bigr)
                        \eta_{D_0/D, v}\Bigr)
\end{equation}
that converges in the limit $D \to \infty$ to
\begin{equation}
  \label{eq:eesee}
  \lim_{D \to \infty} h_x(u,\ep{-t S\up{0,\bullet}_{D,f}}(x,y)v)
      = (\eta_{0, u\,}, \ep{-t T_f} K_f \eta_{0, v}) \, .
\end{equation}
This can be deduced from the strong continuity of $\eta_{c,w}$ 
in $c$ for any $w \in \L$ and the
strong convergence stated in Consequence~\ref{prop:sgcv} together with the
uniform boundedness (according to the Banach-Steinhaus theorem)  
of  
$\exp\bigl(-t S\up{0,\bullet}_{D - 2 D_0,(D - 2 D_0)f/D}\bigr)$ in $D > 2 D_0$.

To finish the proof, we observe that the right-hand side of \eq{eesee} 
is an integral kernel for $\exp(-t T_f)K_f$ on $L_\hol^2(hm)$
that is, in addition, continuous in $x$ and $y$ 
and therefore coincides with the right-hand side of \eq{etatok}.
The continuity of \eq{eesee} is guaranteed by the continuity of the heat kernel
derived in Appendix~\ref{app:hk}, 
and with $K_f \exp(- S\up{0,\bullet}_{D_0,0}) = K_f$
it can be checked that it indeed constitutes an integral kernel.
\end{proof}

\begin{remarks}
The probabilistic representation of Berezin-Toeplitz semi\-groups according to
formula~\eq{DKformula} has also been called a Wiener-regularized
path integral, because it gives meaning to similar, non-rigorous versions of 
such path integrals.

With particular choices of holomorphic line bundles over homogeneous K\"ahler
manifolds related to Lie-group representations \cite{Ono75}, 
formula \eq{DKformula} yields an analogue of the situations considered 
by Daubechies, Klauder, and Paul \cite{DK85,DKP87}. The particular
Lie groups in consideration are the Heisenberg-Weyl group,
$SU(2)$ and the affine group. In each of these cases, the complex 
dimension of the manifold $\M$ is $n=1$, the Riemannian metric $g$
is the real part of the K\"ahler metric, and the imaginary,
skew-symmetric part is in the prequantum 
relation~\eq{pqrel} with the curvature of the line bundle. 
For an explicit result that does not satisfy this relation,
see the treatment in \cite{BLW99} or \cite{BLW99b}. 
The result given there differs from that of Daubechies and Klauder 
\cite{DK85} by a conformal rescaling of the K\"ahler metric on the 
base manifold.

It is worth pointing out that with a suitable
analyticity argument, one could obtain from formula \eq{DKsgform} the probabilistic
expression for the Schwartz kernel of the unitary group $e^{-i t T_f}$, 
which was a primary motivation for \cite{DK82,DK85,DK86,DKP87,KD83,KD84}. 
The case of bounded $f$ may be treated according to \cite{BLW99}. The
techniques in \cite{Wit00} appear suitable for a generalization to 
$f \in \Kato$, but the Kato decomposable case seems to require an additional effort.
\end{remarks}

% -------------------------------
\section{Summary and Outlook}
\label{ch:7}
% -------------------------------

In this work, we have studied a 
coordinate-independent quantization prescription
in the spirit of Berezin and its
representation by Wiener-regularized
path integrals according to an idea of
Daubechies and Klauder. In the present version, 
these path integrals express semigroups 
that are generated by self-adjoint semibounded 
Berezin-Toeplitz operators on 
a generalized Bergman space.

The first results concerned conditions that 
guarantee self-adjointness and semiboundedness 
of Berezin-Toeplitz operators. The use of quadratic 
forms provided a convenient framework to develop 
such conditions, which in the course of 
Sections~\ref{ch:3} to \ref{ch:5} evolved from rather 
abstract form-boundedness to the more concrete 
requirement in terms of the Kato class. The 
Dirichlet Laplacian provided a natural
geometric characterization of this class.
Besides the Kato class, 
the holomorphic Laplacian proved central to our
implementation of the concept by Daubechies
and Klauder on K\"ahler manifolds.
More specifically, we considered 
perturbations of the holomorphic Laplacian
in conjunction with a limiting procedure and
the Feynman-Kac formula to construct
Wiener-regularized path integrals,
a probabilistic representation
of the Schwartz kernel for the semigroup 
generated by a Berezin-Toeplitz operator.
One implication of this construction was that
the reproducing kernel of a space of holomorphic, 
square-integrable sections in a holomorphic Hermitian 
line bundle over a K\"ahler manifold
could be expressed in  purely geometric terms. 
%To this end, we let the K\"ahler metric govern a Brownian
%motion on the base manifold 
%and used the connection of the bundle to lift the process 
%horizontally into the fibers. In the ultra-diffusive limit,
%the expectation of the reverse horizontal transport
%with respect to the conditional Brownian
%motion gave the desired reproducing kernel.
%A similar coordinate-independent expression resulted for the integral
%kernel of a Berezin-Toeplitz semigroup.

The fundamental idea behind all those results was
the relation between Be\-re\-zin-Toeplitz operators and 
Schr\"o\-din\-ger operators, which enabled us to transfer all the relevant
analytic and probabilistic techniques. 

As to further developments,
one may ask whether Wiener-regularized path integrals can also be
found for continuous representations without underlying complex structures.
A step in this direction this has been pointed out by \cite{AK96}
with the use of Dirac operators and spin$^c$ structures. 
Indeed, the completeness argument for the Hilbert space
in Appendix~\ref{app:CTB}
could be applied to a space of merely harmonic functions, since all that
is required are mean-value and continuity properties.
In addition, the context of Dirac operators
may provide enough analytic tools to replace
techniques that so far relied on 
the presence of complex structures.
Another ramification is the concept of path transformations,
well-known in the study of Schr\"odinger operators \cite{Wit00}. 
Indeed, one may use an invariance property of Brownian motion
under harmonic morphisms to relate the resolvents of different
Berezin-Toeplitz operators to one another \cite{Bod}. 
Finally, it may be worthwhile to study the use of Wiener-regularized
path integrals to extend the correspondence principle from the
compact K\"ahler case to non-compact manifolds. 
Probabilistic representations have often been useful to bridge between 
different function spaces. In this case, a suitable procedure
of approximating non-compact K\"ahler manifolds by compact ones
in the path-integral representation
could help enlarging the validity of the correspondence principle.

% non-square integrable
% no holomorphic torsion Dirac etc

%\backmatter

\paragraph{Acknowledgements}
Heartfelt thanks go to John Klauder whose strong intuition
proved invaluable in our discussions of geometric principles 
in quantization procedures.

\begin{appendix}

\section{Completeness of the Generalized Bergman Space} \label{app:CTB}
\label{app:A}
In this part of the appendix, we show that the spaces of holomorphic, square-integrable
sections $\Bg$ we consider are indeed Hilbert spaces. The main part is a localization
argument that reduces the setting to that of the space of square-integrable 
holomorphic functions on the unit ball as studied by Bergman \cite{Ber70}.

Let $\dz$ denote the Lebesgue measure on $\C^n$.
We recall that the open \mbox{ball} 
$B(x,r):=\{ y \in \C^n : \sum_{l=1}^n \abs{y_l-x_l}^2 < r^2 \}$
centered at $x \in \C^n$ with radius $r \ge 0$ has the volume
$\int_{\C^n} \chi_{B(x,r)}(z)\, \dz = \pi^n r^{2n} / n!$,
where $n! := 1 \cdot 2 \cdots n$ denotes the factorial of $n$.

\begin{applem}
The inner-product space $L^2_{\hol}(B(0,1))$ of holomorphic functions
that are square-integrable with respect to the Lebesgue measure on $B(0,1)$
is complete in the norm-topology induced by the usual $L^2$-inner product. 
\end{applem}
\begin{proof}
  Let $(f_j)_{j \in \N}$ be a Cauchy sequence in $L^2_{\hol}(B(0,1))$. First we show 
uniform convergence on all compact sets $C$ inside $B(0,1)$. For any such set $C$,
we can find a nonzero safety radius $r>0$ 
smaller than the distance from $C$ to the boundary of $B(0,1)$. 
Using the mean value property for holomorphic functions,
we can express the difference of two function values at a point
$x \in C$ as the difference between the averages of the two functions,
each computed for a disk centered at $x$.
Then Jensen's inequality in conjunction with the convexity of 
the square-modulus function 
$c \mapsto \abs{c}^2$ on $\C$ yields 
\begin{gather}
  \label{eq:unifcvg}
 \sup_{x \in C} \abs{f_j(x) - f_k(x)}^2 
 %     &=&  \left(\frac{n!}{\pi^n r^{2n}}\right)^2 \sup_{x \in C} \Bigl\vert 
 %          \int_{B(x,r)}  (f_j(z) - f_k(z)) \, \dz \Bigr\vert^2  \\
   \le  \left(\frac{n!}{\pi^n r^{2n}}\right)^2 \sup_{x \in C} \int_{B(x,r)}  
                              \abs{f_j(z) - f_k(z)}^2 \, \dz \\
   \le    \left(\frac{n!}{\pi^n r^{2n}}\right)^2 \int_{B(0,1)}   
                                 \abs{f_j(z) - f_k(z)}^2 \, \dz 
   = \left(\frac{n!}{\pi^n r^{2n}}\right)^2 \norm{f_j - f_k}_2^2 \, .
\end{gather}
The right-hand side can be made arbitrarily small and thus the sequence 
$(f_j)_{j \in \N}$ converges uniformly on $C$. 
By a standard argument in complex analysis, 
we conclude that the pointwise limit defines a holomorphic function 
$f:$ $f(z) =$ $\lim_{j \to \infty} f_j (z)$ in $B(0,1)$.

That the convergence $f_j \to f$ is also in the 
sense of the $L^2$-norm follows from the Cauchy property
of the sequence and from
$
  \norm{f - f_k}_2 \leq$ 
$ \liminf_{j \to \infty} \norm{f_j - f_k}_2 
$
due to pointwise convergence and Fatou's lemma.
\end{proof}

\begin{appthm}
  Let $\L$ be a holomorphic line bundle with the complex, $n$-dimensional 
base manifold $\M$. Suppose the fibers of $\L$ are
equipped with a Hermitian metric $h$ and $M$ possesses
a volume form $m$, both forms non-degenerate and smooth.
Then the space $\Bg$ of holomorphic, square-integrable sections in $\L$
forms a Hilbert space.
\end{appthm}
\begin{proof}
To begin with, we choose an atlas of local trivializations 
$\{\xi_j\}_{j \in I}$
and corresponding reference sections $s_j$. That is, in each chart
domain $U_j$ of the underlying atlas covering $\M$ we choose 
$s_j$ such that the composition $\xi_j \circ s_j$ has the constant
value one in the second component. Choosing $\xi_j: \pi^{-1}(U_j) \to V_j \times \C$
determines $s_j$. 
Thus, we can identify each section $\psi$ with a set of functions 
$\{\psi_j: U_j \to \C\}_{j \in I}$ that satisfy $\psi\vert_{U_j}=\psi_j s_j$
via multiplication in the fibers. 

Similarly as in the preceding completeness proof, it is enough
to show that any given Cauchy sequence 
converges pointwise to
a holomorphic section, which is a holomorphic 
representative of the limit in the $L^2$-sense. 

To this end, we note that with the local reference sections,
the Cauchy sequence $\{\psi^{(l)}\}_{l \in \N}$ is represented 
by a sequence of holomorphic functions $\psi^{(l)}_j: U_j \to \C$. 
The image measure of $\meas$ under a chart $\phi_j: U_j \to V_j \subset \C^n$ 
has a density with respect to the Lebesgue measure on $V_j$, $dm(z)=m_j(z)\dz$.

Suppose we have chosen local trivializations $\xi_j$
with the range of each underlying chart $\phi_j$ being 
a ball of radius $r_j$ centered at the origin. 
The non-degeneracy and smoothness of $h$ and $\mu$
imply that for each $U_j$, there is a strictly positive
lower bound $0 < \epsilon_j < m_j(z)$. 
We deduce that $\bigl\{\psi^{(l)}_j\bigr\}_{l \in \N}$ is 
a Cauchy sequence of holomorphic functions in the
conventional Bergman space $L^2_{\hol}(V_j, \epsilon_j d^{2n}z)$. 
According to the preceding lemma, the sequence converges pointwise to a 
holomorphic function. The limits
obtained on each $V_j$ can then be recombined with the help of the reference 
sections $s_j$ to give a global, holomorphic section. 
This limit section is the holomorphic representative 
that coincides almost everywhere with the limit of the Cauchy sequence 
$\{\psi\up l\}_{l \in \N}$ taken in $L^2(hm)$. 
\end{proof}

\section{Smoothness of Heat Kernels} \label{app:hk}

The crucial idea used in the construction of the heat kernel is that
the index $m \in \N$ of a Sobolev space $W^{m,2}(\R^{\mkern0.5mu d})$ controls
the regularity properties of its functions. 

To simplify the notation, we introduce a customary 
$d$-dimensional multi-index $j=(j_1,j_2, \dots j_d)$ 
with non-negative components $j_1, j_2, \dots j_d \in \bZ^+$
and define its 
degree by $\abs j:=\sum_{k=1}^d j_k$.
For $k=(k_1,k_2,\dots,k_d) \in \R^{\mkern0.5mu d}$, we abbreviate 
$k^j:=k_1^{j_1} k_2^{j_2} \dots k_d^{j_d}$.

\begin{appdef}
  The Sobolev space  $W^{m,2}(\R^{\mkern0.5mu d})$ with $m \in \N$ 
consists of square-inte\-gra\-ble functions $f: \R^{\mkern0.5mu d} \to \C$ 
having Fourier transforms 
$\tilde f:$ $k \mapsto \int_{\R^{\mkern0.5mu d}} e^{-ik\cdot x}$ $f(x) d^d x$
that render the Sobolev norm 
$\int_{\R^{\mkern0.5mu d}} |\tilde f(k)|^2(1+k^2)^m d^{\mkern2mu d}k$ finite.
Equipped with this norm, $W^{m,2}(\R^{\mkern0.5mu d})$ is complete.
\end{appdef}

\begin{applem}
Given a fixed maximal degree $0 \leq l < m - d/2$, the linear functionals
\begin{align}
  \label{eq:deltafn}
  \delta^{(j)}_x: f &\longmapsto \int_{\R^{\mkern0.5mu d}} k^j
                         e^{i k \centerdot x} \tilde f (k) \frac{d^{\mkern2mu d} k}{(2\pi)^d}  
\end{align}
with $x \in \R^{\mkern0.5mu d}$ and $\abs j \leq l$
are uniformly bounded on $W^{m,2}(\R^{\mkern0.5mu d})$. Moreover, in this case any function 
$f \in W^{m,2}(\R^{\mkern0.5mu d})$ has an $l$-times continuously differentiable representative
$x \mapsto \delta^{(0)}_x(f)$.
\end{applem}
\begin{proof}
This statement is a rearrangement of \cite[Theorem 12.29]{CFKS87}.
\end{proof}

%\begin{defn}
%  A uniformly elliptic strictly second order operator $A$
%on a riemannian manifold $\M$
%is defined by the following property: Given a compact set $C$,
%there is a strictly positive lower bound $c>0$ for
%the quadratic form belonging to $A$
%when it is restricted to  $\psi \in C_c^\infty(\M)$,
%\begin{equation}
%   (\psi, A \psi) \ge c \norm{\grad \psi}^2 \, .
%\end{equation}
%\end{defn}

\begin{approp}
  Given a complex line bundle $\L$ with a Riemannian base manifold $\M$, 
the semigroup generated by the self-adjoint Bochner-Laplacian $-\Delta^\cL$ as 
defined in \eq{BochnerLaplacian} has a Schwartz kernel
$\{p^\cL_t(x,y): \L_y \to \L_x\}_{t>0;x,y \in \M}$ that is smooth
in the parameters $t$, $x$ and $y$. 
\end{approp}
\begin{proof} As a first step, we establish properties
of point-evaluation functionals on Sobolev-type spaces
of sections in $\L$. 

A section 
$\sigma=\sigma_j s_j$ with compact support in the domain of 
a chart $\phi_j: U_j \to V_j \subset \R^{\mkern0.5mu d}$
can be identified with $\sigma_j \circ \phi_j^{-1}$,
and because of its compact support canonically extends by zero on the 
remaining part of $\R^{\mkern0.5mu d}$. 
Due to the smoothness and non-degeneracy of the metric,
its eigenvalues obtain a maximum and a nonzero minimum
on the support of $\sigma$. 
Therefore, $\Delta^\cL$ acts locally
as a uniformly elliptic operator and allows estimating 
$(\sigma, (1-\Delta^\cL) \sigma)$
from above and below by multiples of the
Sobolev-norm of the function $\sigma_j$ in  $W^{1,2}(\R^{\mkern0.5mu d})$. 
By an inductive procedure, the same technique
gives estimates for $(\sigma, (1-\Delta^\cL)^m\sigma)$
in terms of norms in $W^{m,2}(\R^{\mkern0.5mu d})$. 
From now on, we refer to the Sobolev-type space of sections $\psi$
having the finite norm $\nnorm{\psi}:=\norm{(1-\Delta^\cL)^{m/2}\psi}$ as
$\cW^{m,2}_\cL(\M)$.

In analogy to the Sobolev spaces on $\R^{\mkern0.5mu d}$,
the linear functional 
$\vartheta_u: \psi \mapsto h_x(u,\psi(x))$ 
evaluating sections at $x=\pi(u)$ is
for sufficiently large $m$ 
bounded in $\cW_\cL^{m,2}(\M)$. 
At first, the bound is only valid on the closed subspace of sections
$\sigma$ with support in a sufficiently small compact set $C$ 
containing $x$. However, the sections in the orthogonal complement
of the subspace vanish on $C$ and thus the bound of $\vartheta_u$ 
passes unchanged to the whole of $\cW_\cL^{m,2}(\M)$ \cite{Hed86}.
By a similar localization argument and the preceding lemma,
$u \mapsto \vartheta_u$ is seen to be smooth, and so are all the sections
$\psi$ in $\cW_\cL^{m,2}(\M)$.

The next step of the proof makes use of these smoothness properties
to construct the heat kernel.

In the spectral representation we see that 
for fixed $m \in \N$ and $t_0>0$, the operators
$(1-\Delta^\cL)^{m/2}\ep{t\Delta^\cL}$
are uniformly in $t\ge t_0$ bounded on $L^2(h\mu)$.
In consequence, the semigroup $e^{t\Delta^\cL}$ is bounded
as a mapping from $L^2(h\mu)$ into all Sobolev-type spaces $\cW^{m,2}(\M)$, 
and choosing a sufficiently large $m$ proves that
the functional $\psi \mapsto \vartheta_u(e^{t\Delta^\cL}\psi)$
is bounded and linear in $\psi \in L^2(hm)$.
By the Riesz Representation Theorem and due to the linearity
of $u \mapsto \vartheta_u$ in the fibers, there is a vector 
$q_t(\centerdot,\pi(u))u$ in $L^2(hm)$ such that 
$\vartheta_u(e^{t\Delta^\cL}\psi)=(q_t(\centerdot,\pi(u))u,\psi)$ for all $\psi \in L^2(hm)$.
By the smoothness of $u\mapsto (q_t(\centerdot,\pi(u))u,\psi)$ and a uniform boundedness
argument, the map $u \mapsto q_t(\centerdot,\pi(u))u$ is smooth in the strong sense. 

In addition, the map $t \mapsto q_t(\centerdot,\pi(u))u$ is also smooth, because
$(1-\Delta^\cL)^{m/2}\ep{t\Delta^\cL}$ is real analytic in $t>0$.
 
In the last step, we define a smooth kernel $p^\L_t(x,y)$ by
\begin{equation}
  \label{eq:ptsmooth}
  h_x(u,p^\L_t(\pi(u),\pi(v))v) := (q_{t/2}(\centerdot,\pi(u))u, q_{t/2}(\centerdot,\pi(v))v)
\end{equation}
and claim that it is a Schwartz kernel for $\ep{t\Delta^\cL}$.
Using the definition \eq{ptsmooth}, the equation
\begin{equation}
   \int_{\M \times \M}  h_x( \psi(x), p_{t}(x,y) \sigma(y))  d\mu(x) \, d\mu(y)    
  =  (\ep{-t\Delta^\cL/2} \psi, \ep{-t\Delta^\cL/2} \sigma) 
\end{equation}
follows for sections $\psi,\sigma \in \Cinfty_{c\L}(\M)$. The self-adjointness
and boundedness of $\ep{-t\Delta^\cL/2}$ then completes the proof.
\end{proof}

% ------------------------------------------------------------------------------
\section{A Version of the Feynman-Kac Formula for Perturbations of
         the Bochner Laplacian} \label{app:ItoLB}

This appendix is concerned with a proof of formula \eq{FKII}.
The strategy followed here is a combination of ideas as presented by Simon 
\cite[Chapter V]{Sim79}, Bismut \cite[Chapitre IX]{Bis81}, and Wittich \cite{Wit00}.
The core portion of the proof is a version of It\^o's formula for sections in 
line bundles, which will be derived first. The remaining part is an approximation argument.

We will use the same notation as in the main text, so $\L$ is a Hermitian line
bundle with a connection $\nabla$ and an associated metric-preserving horizontal
transport $H$. The $d$-dimensional base manifold $\M$ is complete with 
respect to the topology induced by a Riemannian metric. As usual, the 
Brownian motion
in $\M$ with the diffusion constant $D>0$ and the starting point $x$ is 
denoted by $\sB$, and the underlying probability measure 
by $\bP^D_x$. Moreover, $\M$ is assumed to be Brownian-complete and
its Ricci curvature bounded from below.

The Bochner Laplacian is denoted by $\Delta^\cL$.
An additive perturbation to $-D\Delta^\cL$
by a function $q$ as discussed in Definition~\ref{def:Schrop} results
in the Schr\"odinger operator $S_{D,q}^\cL$.
At first, we focus on the unperturbed case.

\begin{applem} \label{lem:shtlocal}
Given a smooth section $\psi$ in $\L$, then for $t\ge 0$
\begin{equation} \label{eq:sFTC}
  H^{-1}_{\sB,t} \psi(\sB_t) = \psi(\sB_0)
 + \sum_{k=1}^d\int_0^t H^{-1}_{\sB,r} \nabla_{E_k} 
             \psi(\sB_r) \langle E_k^\flat,\delta\sB\rangle_r
\end{equation}
relates the inverse of the horizontal transport 
$H_{\sB,\cdot}$ along $\sB$ and the connection $\nabla$ in a 
Stratonovich-integral
equation. As usual, the right-hand side is invariant with respect to the particular
choice of the section $\{E_k\}_{k=1}^d$ in a local orthonormal frame bundle.
The brackets $\langle\cdot,\cdot\rangle$ denote a dual pairing, here with 
the one-form $E_k^\flat=g(E_k, \cdot)$.
\end{applem}
\begin{proof}
By localization \cite{Sch80}, it suffices to check this on a stochastic 
interval $[\mkern-3mu[0,\tau]\mkern-3mu]$, 
where $\tau$ is the exit time of $\sB$ from the
chart domain $U_j$ containing the starting point $\sB_0$.
The proof is accomplished using a local formulation 
of horizontal transport. To this end, we select 
a local trivialization $\xi_j$ and reference section $s_j$ around the $\sB_0$
and associate with each section $\psi$ the representing function
$\psi_j$ satisfying $\psi|_{U_j}=\psi_j s_j$. The so-called connection
one-form determined by $\nabla$ has a local representative $\alpha_j$
that satisfies $\nabla_X(\psi_j s_j)= (X(\psi_j)-i \alpha_j(X) \psi_j) s_j$
for all smooth $\psi$ and vector fields $X \in \Upsilon(\M)$. 

To simplify the notation, we define semimartingales $\sY$ and $\sZ$ 
on $[\mkern-3mu[0,\tau]\mkern-3mu]$ by
\begin{align}
  \sY_t :=\psi_j(\sB_t)
\text{\ \  and\ \  }
  \sZ_t :=\ep{-i\int_0^t\langle \alpha_j,\delta \sB \rangle} \, ,
\end{align}
and use the shorthand $\delta\sW\up{k} = \langle E^\flat_k, \delta \sB \rangle$,
which represents the components of a Brownian motion $\sW$ in $\R^{\mkern0.5mu d}$
that is restricted to the stochastic interval.
In conjunction with Stratonovich stochastic integrals, an
integration by parts rule applies, 
\begin{align}
  \sZ_t \sY_t - \sY_0 &= \int_0^t \sZ_r \delta \sY_r + \int_0^t \sY_r \delta \sZ_r \\
     &= \sum_{k=1}^d \int_0^t \sZ_r E_k(\psi_j)(\sB_r)\delta\sW\up{k}_r
                   - i \sum_{k=1}^d \int_0^t \sZ_r \sY_r \alpha_j(E_k)(\sB_r)\delta \sW\up{k}_r\\
    &= \sum_{k=1}^d \int_0^t \sZ_r \bigl(E_k(\psi_j)
                                          -i \alpha_j(E_k)\psi_j
                                   \bigr)(\sB_r)\delta\sW\up{k}_r \, , \label{eq:Itosdream}
\end{align}
and after reinserting the definitions of $\sY$
and $\sZ$, we obtain an identity which,
together with the localized expression for reverse
horizontal transport, shows that both sides of equation \eq{sFTC}
are the same scalars multiplying $s_j(\sB_0)$.
\end{proof}

\begin{approp}
With the same notation as in the preceding lemma,
a version of the It\^o formula in fiber bundles
is expressed as
\begin{equation} \label{eq:MalliavinIIrep}
  H^{-1}_{B,t} \psi(\sB_t) = \psi(\sB_0) + 
  \sum_{k=1}^d \int_0^t H^{-1}_{\sB,r} \nabla_{E_k} \psi(\sB_r) 
                                             d \sW^{(k)}_r 
 + \int_0^t H^{-1}_{\sB,r} D \Delta^\cL \psi(\sB_r) dr \, .
\end{equation}
\end{approp}
\begin{proof}
As the first step of the proof, we repeat the calculation in the preceding lemma, with
$\sY_t$ replaced by $\sY_t\up{k}:=\bigl(E_k(\psi_j)-i\alpha_j(E_k)\psi_j\bigr)(\sB_t)$, 
which yields
\begin{align}
  \sZ_t \sY\up{k}_t -\sY\up{k}_0 &= \sum_{l=1}^d \int_0^t \sZ_r 
                         \Bigl(    E_l (E_k(\psi_j)) - (\Cov_l E_k)\psi_j
                                  -i E_l(\alpha_j(E_k)) -
                                                              \Bigr. \nonumber \\
                      & \Bigl. \phantom{= \sum_{l=1}^d}   
                   i \alpha_j(\Cov_l E_k) 
                        -i \alpha_j(E_l)(E_k(\psi_j)-i\alpha_j(E_k)\psi_j) \Bigr)
                         (\sB_r) \delta \sW\up{l}_r \, . \label{eq:messymessy}
\end{align}
The covariant derivative of the frame vectors enters because those are not
horizontally transported along $\sB$.

Now we convert equations \eq{Itosdream} and \eq{messymessy}
to  It\^o differentials
and insert the stochastic integral expression  for $\sZ \sY\up{k}$
into the cross variation emerging from equation~\eq{Itosdream},
\begin{align}
  \sZ_t \sY_t - \sY_0 &= \sum_{k=1}^d \int_0^t \sZ_r \sY\up{k}_r \delta\sW\up{k}
                           + \frac{1}{2} \sum_{k=1}^d \qvar{ \sZ \sY\up{k} , \sW\up{k} }_t\\
            &= \sum_{k=1}^d  \int \sZ_r \sY\up{k}_r d\sW\up{k} 
                  + \frac{1}{2} \sum_{k,l=1}^d \int_0^t \sZ_r 
  \Bigl( \bigl(   E_l (E_k(\psi_j)) - (\Cov_l E_k)\psi_j \bigr.\Bigr.\nonumber\\
                         &\phantom{=} \Bigl.\bigl.
           -i E_l(\alpha_j(E_k)) - i \alpha_j(\Cov_l E_k) \bigr) (\sB_r)
                        -i \alpha_j(E_l) \sY\up{k}_r\Bigr) d \qvar{\sW\up{l},\sW\up{k}}_r \, .
\end{align}
After contracting the summation indices with the cross variation
$\qvar{\sW\up{l},\sW\up{k}}_r=2D\delta_{lk}r$,
a similar identification as in the preceding lemma 
and the differential operator expression obtained for $\Delta^\cL$
proves formula \eq{MalliavinIIrep}.
\end{proof}

\begin{appcon}
If $\psi \in L^2(hm)$ and $\E^D_x[\bullet]$ denotes the expectation with
respect to the Brownian motion starting at $x \in \M$, then 
for $t \ge 0$ the semigroup
generated by $D\Delta^\cL$ can be represented as 
\begin{equation} \label{eq:BLSG}
  e^{tD\Delta^\cL} \psi (x) = \E^D_x[ H_{\sB,t}^{-1}\psi(\sB_t) ]
\end{equation}
\end{appcon}
\begin{proof}
  First, we assume $\psi \in \Cinfty_{c\L}(\M)$ and abbreviate
$P^\cL_{D,t} \psi(x):=\E_x(H_{\sB,t}^{-1}\psi(\sB_t)]$.
Since $H^{-1}_{\sB,t}$ preserves the Hermitian metric $h$, each $P^\cL_{D,t}$
is seen to be a bounded operator. Moreover, by the time reversal
invariance of Brownian motion it is self-adjoint. Finally, the
family $\{P^\cL_{D,t}\}_{t \ge 0}$ forms a semigroup due to the Markov property
\begin{align}
  P^\cL_{D,t+s} \psi(x) &= \E^D_x[ H_{\sB,t+s}^{-1}\psi(\sB_{t+s}) ]\\
                  &= \E^D_x[ H_{\sB,t}^{-1} \E^D_{\sB_t}[ H_{\sB,s}^{-1}\psi(\sB_s)] ]
                  = P^\cL_{D,t}(P^\cL_{D,s} \psi)(x) 
\end{align}
valid for $s,t \ge 0$.
To verify that both sides of \eq{BLSG} are identical, we note that
the generators agree on $\psi \in \Cinfty_{c\L}(\M)$, because $P^\cL_{D,t}\psi$ 
satisfies the same integral equation as $e^{tD\Delta^\cL}\psi$,
\begin{align}
  P^\cL_{D,t} \psi(x) &= \E^D_x[H_{\sB,t}^{-1} \psi(\sB_t)]\\
              &=  \E_x^D\Bigl[\psi(\sB_0)+\int_0^t H^{-1}_{\sB,s} D \Delta^\cL \psi(\sB_s) ds \Bigr]\\ 
              &= \psi(x) + \int_0^t P^\cL_{D,s} D\Delta^\cL \psi(x) ds \, .
\end{align}
By the definition of $P^\cL_{D,t}$, the semigroup can be defined on 
all $\psi \in L^2(hm)$.
Therefore, its generator defines a self-adjoint
extension of $D\Delta^\cL|_{\Cinfty_{c\L}(\M)}$, but this is necessarily $D\Delta^\cL$,
because the latter is essentially self-adjoint on $\Cinfty_{c\L}(\M)$.
\end{proof}

\begin{appthm}
  If the assumptions listed at the beginning of this appendix
are satisfied, 
$\psi \in L^2(hm)$, and $q \in \Katopm$, then the semigroup
$e^{-tS^\cL_{D,q}}$ generated by the Schr\"odinger operator $S_{D,q}^\cL$ 
has the probabilistic representation
\begin{equation}
  e^{-tS_{D,q}^\cL}\psi(x) = \E^D_x\left[ e^{-\int_0^t q(\sB_s)ds} H_{\sB,t}^{-1}\psi(\sB_t)\right] \, ,
\end{equation}
valid for $m$-almost every $x \in \M$.
\end{appthm}
\begin{proof}
  First, we suppose $q$ is continuous, $\psi$ is a smooth section, and
both are bounded. Then, along the lines of \eq{MalliavinIIrep} and with the integration
by parts rule,
\begin{align}
  e^{-\int_0^t q(\sB_s)ds} H_{\sB,t}^{-1} \psi(\sB_t)
 &= \psi(\sB_0) + \sum_{k=1}^d \int_0^t e^{-\int_0^r q(\sB_s)ds}
                H_{\sB,r}^{-1} \nabla_{E_k} \psi(\sB_r) d\sW\up{k}_r\nonumber \\
 & \phantom{=sum} + \int_0^t e^{-\int_0^r q(\sB_s)ds} H_{\sB,r}^{-1} (D\Delta^\cL \psi
                       - q \psi)    (\sB_r) dr  \, \label{eq:sginst}.
\end{align}
Since $q$ is bounded, the modification of the heat semigroup 
defined by inserting \eq{sginst} in the expectation value of \eq{BLSG}
has as its generator a self-adjoint extension of
$(D\Delta^\cL - q)|_{\Cinfty_{c\L}(\M)}$. Again, by essential self-adjointness,
this is seen to be the difference $-S^\cL_{D,q}=D\Delta^\cL - q$.

In the last step, we approximate the general case $q \in \Katopm$ by truncation.
We define the net $\{q\up{l}_k\}$ of bounded functions
\begin{equation}
  x \mapsto q\up{l}_{k} (x) := \min \{ \max \{ q(x),-k \},l \}
\end{equation}
indexed by $k,l \in \N$. Truncating $q\in \Katopm$ in this manner
gives
\begin{equation}
  e^{-t S^\cL_{\raisebox{0pt}[4pt]{\tiny$D,q\up{l}_{k}$}}}\psi(x) = \E^D_x\left[ e^{-\int_0^t q\up{l}_{k}(\sB_s) ds} 
               H_{\sB,t}^{-1}\psi(\sB_t)\right]
\end{equation}
valid for $m$-almost every $x$ by the above argument. Now, by monotone form convergence
we obtain strong convergence on the left when consecutively first $l \to \infty$ and
then $k \to \infty$, whereas on the right dominated convergence applies to
both limits, because
\begin{equation}
 \E_x^D\left[ e^{\int_0^t q^-(\sB_s) ds} h_{\sB_t}(\psi(\sB_t),\psi(\sB_t))\right] < \infty
\end{equation}
since the negative part $q^- \in \Kato$ and $x \mapsto h_x(\psi(x),\psi(x))$
are $m$-integrable.
\end{proof}
\end{appendix}


\begin{thebibliography}{10}
\expandafter\ifx\csname url\endcsname\relax
  \def\url#1{\texttt{#1}}\fi
\expandafter\ifx\csname urlprefix\endcsname\relax\def\urlprefix{URL }\fi

\bibitem{Ber72a}
F.~A. Berezin, Covariant and contravariant operator symbols, Math. USSR
  Izvestija 6 (1972) 1117--1151, russ. orig.: Izv. Akad. Nauk SSSR Ser. Mat.
  {\bf 36} (1972), 1134--1167.

\bibitem{Ber74}
F.~A. Berezin, Quantization, Math. USSR Izvestija 8 (1974) 1109--1165, russ.
  orig.: Izv. Akad. Nauk SSSR, Ser. Mat. {\bf 38} (1974), 1116--1175.

\bibitem{Kla63a}
J.~R. Klauder, Continuous-representation theory. {I}. {P}ostulates of
  continuous-representation theory, J. Math. Phys. 4 (1963) 1055--1058.

\bibitem{Kla63b}
J.~R. Klauder, Continuous-representation theory. {I}{I}. {G}eneralized relation
  between quantum and classical dynamics, J. Math. Phys. 4 (1963) 1058--1073.

\bibitem{Kla64}
J.~R. Klauder, Continuous-representation theory. {I}{I}{I}. {O}n functional
  quantization of classical systems, J. Math. Phys. 5 (1964) 177--187.

\bibitem{KM65}
J.~R. Klauder, J.~McKenna, Continuous-representation theory. {V}.
  {C}onstruction of a class of scalar boson field continuous representations,
  J. Math. Phys. 6 (1965) 68--87.

\bibitem{KMK65}
J.~R. Klauder, J.~McKenna, D.~G. Currie, On ``diagonal'' coherent-state
  representations for quantum-mechanical density matrices, J. Math. Phys. 6
  (1965) 734--739.

\bibitem{MK64}
J.~McKenna, J.~R. Klauder, Continuous-representation theory. {I}{V}.
  {S}tructure of a class of function spaces arising from quantum mechanics, J.
  Math. Phys. 5 (1964) 878--896.

\bibitem{BMS94}
M.~Bordemann, E.~Meinrenken, M.~Schlichenmaier, Toeplitz quantization of
  {K}{\"a}hler manifolds and $gl(n)$, $n \rightarrow \infty$ limits, Commun.
  Math. Phys. 165 (1994) 281--296.

\bibitem{CGR90}
M.~Cahen, S.~Gutt, J.~Rawnsley, Quantization of {K}{\"a}hler manifolds. {I}.
  {G}eometric interpretation of {B}erezin's quantization, J. Geom. Phys. 7
  (1990) 45--62.

\bibitem{CGR93}
M.~Cahen, S.~Gutt, J.~Rawnsley, Quantization of {K}{\"a}hler manifolds. {II},
  Trans. Amer. Math. Soc. 337 (1993) 73--98.

\bibitem{CGR94}
M.~Cahen, S.~Gutt, J.~Rawnsley, Quantization of {K}{\"a}hler manifolds. {III},
  Lett. Math. Phys. 30 (1994) 291--305.

\bibitem{CGR95}
M.~Cahen, S.~Gutt, J.~Rawnsley, Quantization of {K}{\"a}hler manifolds. {IV},
  Lett. Math. Phys. 34 (1995) 159--168.

\bibitem{Kos70}
B.~Kostant, Quantization and unitary representations. {I}. {P}requantization,
  in: Lectures in modern analysis and applications, III, Springer, Berlin,
  1970, pp. 87--208. Lecture Notes in Math., Vol. 170.

\bibitem{Sni80}
J.~{\'S}niatycki, Geometric quantization and quantum mechanics, Springer, New
  York, 1980.

\bibitem{Sou66}
J.-M. Souriau, Quantification g\'eom\'etrique, Comm. Math. Phys. 1 (1966)
  374--398.

\bibitem{Per86}
A.~Perelomov, Generalized coherent states and their applications, Texts and
  Monographs in Physics, Springer, Berlin, 1986.

\bibitem{Sch98}
M.~Schlichenmaier, {B}erezin-{T}oeplitz quantization of compact {K\"{a}}hler
  manifolds, in: A.~Strasburger, S.~Ali, J.-P. Antoine, J.-P. Gazeau,
  A.~Odzijewicz (Eds.), Quantization, Coherent States and Poisson Structures,
  Proceedings of the XIV'th Workshop on Geometric Methods in Physics, Bia\l
  owie\.za, 1995, Polish Scientific Publisher PWN, 1998, pp. 101--115.

\bibitem{DK82}
I.~Daubechies, J.~R. Klauder, Constructing measures for path integrals, J.
  Math. Phys. 23 (1982) 1806--1822.

\bibitem{DK85}
I.~Daubechies, J.~R. Klauder, Quantum-mechanical path integrals with {W}iener
  measure for all polynomial {H}amiltonians. {I}{I}, J. Math. Phys. 26 (1985)
  2239--2256.

\bibitem{DK86}
I.~Daubechies, J.~R. Klauder, True measures for real time path integrals, in:
  M.~L. Gutzwiller, A.~Inomata, J.~R. Klauder, L.~Streit (Eds.), Path Integrals
  from me{V} to {M}e{V}, Bielefeld Encounters in Physics and Mathematics, World
  Scientific, Singapore, 1986, pp. 425--432.

\bibitem{KD83}
J.~R. Klauder, I.~Daubechies, Measures for path integrals, Phys. Rev. Lett. 48
  (1982) 117--120.

\bibitem{KD84}
J.~R. Klauder, I.~Daubechies, Quantum mechanical path integrals with {W}iener
  measures for all polynomial {H}amiltonians, Phys. Rev. Lett. 52 (1984)
  1161--1164.

\bibitem{DKP87}
I.~Daubechies, J.~R. Klauder, T.~Paul, Wiener measures for path integrals with
  affine kinematic variables, J. Math. Phys. 28 (1987) 85--102.

\bibitem{AK96}
R.~Alicki, J.~R. Klauder, Quantization of systems with a general phase space
  equipped with a {R}iemannian metric, J. Phys. A 29 (1996) 2475--2483.

\bibitem{AKL93}
R.~Alicki, J.~R. Klauder, J.~Lewandowski, Landau-level ground state and its
  relevance for a general quantization procedure, Phys. Rev. A 48 (1993)
  2538--2548.

\bibitem{Kla94}
J.~R. Klauder, Quantization on non-homogeneous manifolds, Int. J. Theor. Phys.
  33 (1994) 509--522.

\bibitem{KO89}
J.~R. Klauder, E.~Onofri, Landau levels and geometric quantization, Int. J.
  Mod. Phys. 4 (1989) 3939--3949.

\bibitem{Cha99}
L.~Charles, Feynman path integral and {T}oeplitz quantization, Helv. Phys. Acta
  72 (1999) 341--355.

\bibitem{GH78}
P.~Griffiths, J.~Harris, Principles of algebraic geometry, Wiley, New York,
  1978.

\bibitem{Zha00}
F.~Zhang, Complex Differential Geometry, Studies in Advanced Mathematics, AMS
  and International Press, Providence (R. I.), 2000.

\bibitem{Sim71}
B.~Simon, Quantum mechanics for {H}amiltonians defined as quadratic forms,
  Princeton University Press, Princeton, N. J., 1971, princeton Series in
  Physics.

\bibitem{RS75}
M.~Reed, B.~Simon, Methods of Modern Mathematical Physics, Vol. II, Fourier
  analysis, self-adjointness, Academic Press, New York, 1975.

\bibitem{Kat55}
T.~Kato, Quadratic forms in {H}ilbert spaces and asymptotic perturbation
  series, Department of Mathematics, University of California, Berkeley,
  Calif., 1955.

\bibitem{LM54}
P.~D. Lax, A.~N. Milgram, Parabolic equations, in: Contributions to the theory
  of partial differential equations, no.~33 in Ann. of Math. Stud., Princeton
  University Press, Princeton, N. J., 1954, pp. 167--190.

\bibitem{Lio61}
J.-L. Lions, {\'E}quations diff\'erentielles op\'erationnelles et probl\`emes
  aux limites, Vol. 111 of Die Grundlehren der mathmematischen Wissenschaften,
  Springer, Berlin, 1961.

\bibitem{Nel64b}
E.~Nelson, Interaction of nonrelativistic particles with a quantized scalar
  field, J. Math. Phys. 5 (1964) 1190--1197.

\bibitem{RS80}
M.~Reed, B.~Simon, Methods of modern mathematical physics, Vol. I, Functional
  analysis, Academic Press, New York, 1980.

\bibitem{Wei80}
J.~Weidmann, Linear Operators in {H}ilbert spaces, Vol.~68 of Graduate Texts in
  Mathematics, Springer, New York, 1980.

\bibitem{Cic96}
D.~Cicho\'n, Notes on unbounded {T}oeplitz operators in {S}egal-{B}argmann
  spaces, Ann. Polon. Math. 64 (1996) 227--235.

\bibitem{JS94}
J.~Janas, J.~Stochel, Unbounded {T}oeplitz operators in the {S}egal-{B}argmann
  space, {II}, J. Funct. Anal. 126 (1994) 419--447.

\bibitem{BU96}
D.~Borthwick, A.~Uribe, Almost complex structures and geometric quantization,
  Math. Res. Lett. 3 (1996) 845--861.

\bibitem{Arn89}
V.~I. Arnold, Mathematical methods of classical mechanics, 2nd Edition, no.~60
  in Graduate Texts in Mathematics, Springer, Berlin, 1989.

\bibitem{GW79}
R.~E. Greene, H.~Wu, ${C}^\infty$-approximations of convex, subharmonic, and
  plurisubharmonic functions, Ann. Scient. {\'E}c. Norm. Sup. 12 (1979) 47--84.

\bibitem{Dav89}
E.~B. Davies, Heat kernels and spectral theory, Cambridge Tracts in
  Mathematics, Cambridge University Press, Cambridge, 1989.

\bibitem{Stu92}
K.-T.\ Sturm, Heat kernel bounds on manifolds, Math. Ann. 292 (1992)
149--162.

\bibitem{BGV92}
N.~Berline, E.~Getzler, M.~Vergne, Heat kernels and {D}irac operators, no. 298
  in Grundlehren der mathematischen {W}issenschaften, Springer, 1992.

\bibitem{Sim78}
B.~Simon, A canonical decomposition for quadratic forms with applications to
  monotone convergence, J. Funct. Anal. 28 (1978) 377--385.

\bibitem{Kha59}
R.~Z. Kha\'sminskii, On positive solutions of the equation ${A}u+{V}u=0$,
  Theoret. Probab. Appl. 4 (1959) 309--318.

\bibitem{Szn98}
A.-S. Sznitman, Brownian Motion, Obstacles and Random Media, Springer
  Monographs in Mathematics, Springer, Berlin, 1998.

\bibitem{Sch80}
L.~Schwartz, Semi-martingales sur des vari{\'e}t{\'e}s, et martingales
  conformes sur des vari{\'e}t{\'e}s analytiques complexes, no. 780 in Lecture
  Notes in Mathematics, Springer, Berlin, 1980.

\bibitem{KS91}
I.~Karatzas, S.~Shreve, Brownian motion and Stochastic Calculus, 2nd Edition,
  no. 113 in Graduate Texts in Mathematics, Springer, New York, 1991.

\bibitem{Par67}
K.~R. Parthasarathy, Probability measures on metric spaces, Academic Press, New
  York, 1967.

\bibitem{Dav88}
E.~B. Davies, Gaussian upper bounds for the heat kernel of some second-order
  operators on {R}iemannian manifolds, J. Funct. Anal. 80 (1988) 16--32.

\bibitem{BLW99}
B.~Bodmann, H.~Leschke, S.~Warzel, A rigorous path integral for quantum spin
  using flat-space {W}iener regularization, J. Math. Phys. 40 (1999)
  2549--2559.

\bibitem{Ono75}
E.~Onofri, A note on coherent state representations of {L}ie groups, J. Math.
  Phys. 16 (1975) 1087--1089.

\bibitem{BLW99b}
B.~Bodmann, H.~Leschke, S.~Warzel, A rigorous path-integral formula for quantum
  spin via planar {B}rownian motion, in: R.~Casalbuoni, R.~Giachetti,
  V.~Tognetti, R.~Vaia, P.~Verrucchi (Eds.), Path Integrals from peV to TeV,
  World Scientific, Singapore, 1999, pp. 173--176.

\bibitem{Wit00}
O.~Wittich, A transformation of a {F}eynman-{K}ac formula for holomorphic
  families of type {B}, J. Math. Phys.  (2000) 244--259.

\bibitem{Bod}
B.~G. Bodmann, Relating resolvents of {B}erezin-{T}oeplitz operators by an
  invariance property of {B}rownian motion, in preparation.

\bibitem{Ber70}
S.~Bergman, The kernel function and conformal mapping, 2nd Edition, no.~5 in
  Amer. Math. Soc. Survey, AMS, Providence (R. I.), 1970.

\bibitem{CFKS87}
H.~L. Cycon, R.~G. Froese, W.~Kirsch, B.~Simon, Schr\"odinger operators, with
  application to quantum mechanics and global geometry, Texts and Monographs in
  Physics, Springer, Berlin, 1987.

\bibitem{Hed86}
L.~I. Hedberg, Approximation in {S}obolev spaces and nonlinear potential
  theory, in: Nonlinear functional analysis and its applications, Part 1
  (Berkeley, Calif., 1983), Amer. Math. Soc., Providence, RI, 1986, pp.
  473--480.

\bibitem{Sim79}
B.~Simon, Functional integration and quantum physics, Academic Press, New York,
  1979.

\bibitem{Bis81}
J.-M. Bismut, M{\'e}canique al{\'e}atoire, no. 866 in Lecture Notes in
  Mathematics, Springer, Berlin, 1981.

\end{thebibliography}
\end{document}